\documentclass[fleqn,usenatbib]{mnras}

\usepackage{newtxtext,newtxmath}
\usepackage[T1]{fontenc}

\DeclareRobustCommand{\VAN}[3]{#2}
\let\VANthebibliography\thebibliography
\def\thebibliography{\DeclareRobustCommand{\VAN}[3]{##3}\VANthebibliography}

\usepackage{graphicx}	% Including figure files
\usepackage{amsmath}	% Advanced maths commands
\usepackage{multirow}
\usepackage{cleveref}
\usepackage{tabularx}
\usepackage{arydshln}
\usepackage{pdflscape}
\usepackage{array}

\crefformat{figure}{Fig.~#2#1#3}
\crefformat{equation}{equation~#2#1#3}
\crefformat{section}{section~#2#1#3}
\crefformat{table}{Tab.~#2#1#3}
\crefformat{appendix}{appendix~#2#1#3}
\crefmultiformat{figure}{Figs.~#2#1#3}{~and~#2#1#3}%
    {,~#2#1#3}{,~#2#1#3}
\crefrangeformat{figure}{Figs.~(#3#1#4--#5#2#6)}
\newcolumntype{P}[1]{>{\centering\arraybackslash}p{#1}}
\newcommand{\comment}[1]{}

\title[SARAS2: A Bayesian re-analysis]{A comprehensive Bayesian re-analysis of the SARAS2 data from the Epoch of Reionization}

\author[H. T. J. Bevins et al.]{H. T. J. Bevins,$^{1}$\thanks{E-mail: htjb2@cam.ac.uk}
E. de Lera Acedo$^{1, 2}$,
A. Fialkov$^{2,3}$,
W. J. Handley$^{1, 2}$,
S. Singh$^{4, 5, 6}$,
\newauthor
R. Subrahmanyan$^{7}$,
and R. Barkana$^{8, 9}$
\\
$^{1}$Astrophysics Group, Cavendish Laboratory, J. J. Thomson Avenue, Cambridge, CB3 0HE, UK\\
$^{2}$Kavli Institute for Cosmology, Madingley Road, Cambridge CB3 0HA, UK \\
$^{3}$Institute of Astronomy, University of Cambridge, Madingley Road, Cambridge CB3 0HA, UK \\
$^{4}$ Raman Research Institute, C V Raman Avenue, Sadashivanagar, Bangalore 560080, India \\
$^{5}$ McGill Space Institute, McGill University, 3550 rue University, Montr\'eal, QC H3A 2A7, Canada\\
$^{6}$ Department of Physics, McGill University, 3600 rue University, Montr\'eal, QC H3A 2T8, Canada \\
$^{7}$ Space \& Astronomy CSIRO, 26 Dick Perry Ave, Kensington WA 6151, Australia \\
$^{8}$ School of Physics and Astronomy, Tel-Aviv University, Tel-Aviv, 69978, Israel \\
$^{9}$ Institute for Advanced Study, 1 Einstein Drive, Princeton, New Jersey 08540, USA \\
}

\date{Accepted 2022 April 20. Received 2022 April 1; in original form 2022 January 27}

\pubyear{2022}

\begin{document}
\label{firstpage}
\pagerange{\pageref{firstpage}--\pageref{lastpage}}
\maketitle

\begin{abstract}

We present a Bayesian re-analysis of the sky-averaged 21-cm experimental data from SARAS2 using nested sampling implemented with \textsc{polychord}, spectrally smooth foreground modelling implemented with \textsc{maxsmooth}, detailed systematic modelling and rapid signal emulation with \textsc{globalemu}. Our analysis differs from previous analysis of the SARAS2 data through the use of a full Bayesian framework and separate modelling of the foreground and non-smooth systematics. We use the most up-to-date signal models including Lyman-$\alpha$ and CMB heating parameterised by astrophysical parameters such as star formation efficiency, X-ray heating efficiency, minimal virial circular velocity of star forming galaxies, CMB optical depth and the low energy cutoff of the X-ray spectral energy distribution. We consider models with an excess radio background above the CMB produced via radio emission from early galaxies and parameterised by a radio production efficiency. A non-smooth systematic is identified and modelled as both a frequency damped sinusoid introduced by the electronics and separately from the sky. The latter is modulated by the total efficiency of the antenna and marginally favoured by the data. We consider three different models for the noise in the data. The SARAS2 constraints on individual astrophysical parameters are extremely weak however we identify classes of disfavoured signals. We weakly disfavour standard astrophysical models with high Lyman-$\alpha$ fluxes and weak heating and more confidently disfavour exotic models with high Lyman-$\alpha$ fluxes, low X-ray efficiencies and high radio production efficiencies in early galaxies.

\end{abstract}

\begin{keywords}
reionization -- first stars -- early Universe -- cosmology:observations
\end{keywords}

%%%%%%%%%%%%%%%%%%%%%%%%%%%%%%%%%%%%%%%%%%%%%%%%%%

%%%%%%%%%%%%%%%%% BODY OF PAPER %%%%%%%%%%%%%%%%%%

\section{Introduction}
\label{sec:intro}

The global (sky-averaged) 21-cm signal from Cosmic Dawn~(CD) and the Epoch of Reionization~(EoR) is a theoretically observable average deviation between the spin temperature, $T_\mathrm{s}$, of neutral hydrogen and the radio background, $T_\mathrm{r}$ (typically assumed to be the Cosmic Microwave Background, CMB). The spectral structure of the signal, characterised by an absorption feature and potential emission, can be used to infer information about the large scale structure formation in the early universe, star formation, as well as thermal and ionization histories of the intergalactic medium \citep[IGM, ][]{Furlanetto2006, Pritchard2012, Barkana2016}.

Theoretical models of the global 21-cm signal \citep[e.g.][]{Visbal2012, Mirocha2014, Fialkov2014, Cohen2017, Reis2021} use a set of astrophysical parameters to define the structure of the signal such as the star formation efficiency, $f_*$, the minimal virial circular velocity of star forming galaxies, $V_c$, the X-ray efficiency of sources, $f_X$, the slope of the X-ray spectral energy distribution~(SED), $\alpha$, the low energy cutoff of the X-ray SED, $E_\mathrm{min}$ \footnote{This parameter has previously been referred to in the literature as $\nu_\mathrm{min}$. We make the change in notation here to clarify that this is an energy.}, the mean free path of ionizing photons, $R_\mathrm{mfp}$ and the CMB optical depth, $\tau$. The identification of a global signal and subsequent determination of these parameters is the subject of ongoing experimental work using a variety of different techniques: SARAS3 \citep[Shaped Antenna measurement of the background RAdio Spectrum,][]{Girish2020, SARAS2021, Raghunathan2021}, EDGES \citep[Experiment to Detect the Global Epoch of Reionization Signature,][]{Bowman2018}, REACH~\citep[Radio Experiment for the Analysis of Cosmic Hydrogen,][]{Acedo2019}, PRIZM~\citep[Probing Radio Intensity at High-Z from Marion,][]{Philip2019}, LEDA \citep[Large-aperture Experiment to Detect the Dark Ages,][]{Price2018}, DAPPER (Dark Ages Polarimeter PathfindER, \url{https://www.colorado.edu/project/dark-ages-polarimeter-pathfinder/}) and MIST (Mapper of the IGM Spin Temperature, \url{http://www.physics.mcgill.ca/mist/}) among others.

In 2018 the EDGES collaboration reported the detection of an absorption trough at 78 MHz \citep{Bowman2018}. However, the reported feature is approximately 3 times deeper\footnote{Note that this value is often reported as 2 times deeper and is based on the maximum predicted depth, $\approx 250$ mK of the global signal from simulations like those in \cite{Cohen2017}. However, more recent simulations, including Lyman-$\alpha$ and CMB heating, by \cite{Reis2021} report a maximum predicted depth of $\approx 165$ mK approximately 3 times shallower than the reported $\approx500$ mK signal from EDGES \citep{Bowman2018}.} 
than the current standard theoretical predictions \citep{Reis2021}. While the excess depth can be explained theoretically with an excess radio background above the CMB \citep{Bowman2018, Feng2018, Ewall2018, Jana2018, Mirocha2019, Fialkov2019, Reis2020} or interactions between dark matter and baryons \citep{BarkanaDM2018, Fialkov2018, Barkana2018, Berlin2018, Kovetz2018, Munoz2018, Slatyer2018, Liu2019} there are concerns about the data analysis and potential presence of systematics in the publicly available EDGES data \citep{Hills2018, Bradley2019, Singh2019, Sims2020, Bevins2021}.

Several experiments, both single antenna and interferometers, have provided constraints on the parameter space of the 21-cm signal at redshifts corresponding to the EoR: HERA~ \citep{HERA}, LOFAR ~\citep{10.1093/mnras/staa487, Mondal2020, Greig2021}, MWA ~\citep{Greig2020, 10.1093/mnras/stab776}, EDGES ~\citep{Monsalve_2017, Monsalve_2018, Monsalve2019} and SARAS2 ~\citep{Singh2017, Singh2018}.
We note that the parameterisation and modelling of the signals, as well as the prior ranges, are not always consistent across the literature. However, in general the conclusions disfavour signals with deep absorption features, within the band of each instrument, from inefficient X-ray heating and a sharp reionization feature.

In this paper we present a re-analysis of the SARAS2 data, which targeted the EoR at low redshifts (high frequencies). Previous analysis of 63 hrs of nighttime observations, between October 2016 and July 2017, at the Timbaktu Collective in Southern India with the SARAS2 instrument concluded that scenarios with rapid reionization and weak X-ray heating were disfavoured by the data \citep[][]{Singh2017, Singh2018}. In this analysis the authors used initially a Bayesian likelihood ratio test to determine whether the presence of particular signal models, from a simulated set of 264, were favoured in the data or not \citep{Singh2017}. This was followed by a detailed frequentist approach that ruled out a larger number of simulated signals from the same set of models and using the same data \citep{Singh2018}. Of the tested scenarios 9 were disfavoured by the data in \cite{Singh2017} and 20, of which 15 were rejected with a significance $> 5\sigma$, were rejected in \cite{Singh2018}. There was no reported detection from the analysis. In both cases high order polynomials were used to model the foreground and systematics, in the belief that any present in the data are smooth, in combination. A high level summary of the differences between the analysis in this paper and the previous analysis of the SARAS2 data can be found in \cref{tab:high_level_summary} and these are futher discussed below.

\begin{table*}
    \centering
    \begin{tabular}{|c|p{3cm}|p{4cm}|p{5cm}|}
        \hline
         &\cite{Singh2017} & \cite{Singh2018} & This work \\
         \hline
         \hline
         Analysis Type & Likelihood ratios testing preference of the data for the presence or absence of signals. & Frequentist approach based on that used in \cite{Monsalve_2017}. & Bayesian nested sampling using \textsc{polychord} \citep{Handley2015a, Handley2015b}. \\
         \hline
         Foreground Modelling & \multicolumn{2}{p{8cm}|}{Unconstrained polynomials of varying orders (e.g. N = 4 -8 in \citep{Singh2018}).} & Smooth foreground models based on Maximally Smooth Functions and implemented with \textsc{maxsmooth} \citep{Bevins2021}. \\
         \hline
         Systematic Modelling & \multicolumn{2}{l|}{Assumed to be smooth and modelled with foreground model.} & Identified through use of smooth foreground model and separately modelled with physically motivated functions. \\
         \hline
         Noise Modelling & Derived by accounting for RFI, system temperature, absolute calibration and differences between adjacent channels. & Mock Gaussian distributed noise based on system attributes. & A set of Gaussian models with constant, frequency damped and relative weights based amplitudes.\\
         \hline
         Signal Modelling & \multicolumn{2}{p{8cm}|}{A library of 264 standard astrophysical models with no additional radio background above the CMB~\citep{Cohen2017}.} & Broader study sampling across large prior ranges, for both standard astrophysical models \citep{Reis2021} and exotic astrophysical models with excess radio backgrounds \citep{Reis2020}, using the signal emulator \textsc{globalemu} \citep{Bevins2021b}. \\
         \hline
    \end{tabular}
    \caption{A high level summary of the differences between the previous analysis of the SARAS2 data and the work performed in this paper. The differences are expanded on primarily in \cref{sec:modelling}.}
    \label{tab:high_level_summary}
\end{table*}

Here, we determine parameter constraints over broad prior ranges using the latest astrophysical models of the global 21-cm signal \citep{Reis2020, Reis2021}, representing an improved understanding of the standard astrophysical picture, and the nested sampling algorithm \citep{skilling_nested_2004} \textsc{polychord} \citep{Handley2015a, Handley2015b}. We use models that include Lyman-$\alpha$ heating~\citep{madau_21_1997, chen_lah_2004, furlanetto_scattering_2006, Chuzhoy2007}, CMB heating~\citep{Venumadhav2018} and multiple scattering of Lyman-$\alpha$ photons \citep{semelin_b_lyman-alpha_2007, Naoz2008, baek_s_simulated_2009, visbal_impact_2018, molaro_artist_2019, reis_mapping_2020}. The effects on the global signal of these physical processes have been understood for some time but the magnitude of those effects over a larger parameter space were not understood until recently \citep{Villanueva2020, Mittal2020, Reis2021}. Additionally, we study astrophysical scenarios with a wide range of radio production efficiencies, $f_\mathrm{radio}$, for early galaxies. A subset of the latter models could explain EDGES using an excess radio background. This is the first time that a full Bayesian analysis of data from a global 21-cm experiment has been performed with these specific astrophysical simulations. We note, however, that the value of $f_\mathrm{radio}$ has previously been constrained using the amplitude of the EDGES absorption feature \citep{Reis2020}\footnote{For clarity, note that we assume the astrophysical scenario of enhanced radio emission from galaxies \citep{Reis2020}, and not the more exotic scenario of an external radio background from the dark ages \citep{Fialkov2019}.} and more recently using upper limits on the power spectrum from the Hydrogen Epoch of Reionization Array~\citep[HERA,][]{HERA}.

In this work we use the recently developed emulator \textsc{globalemu} \citep[][]{Bevins2021b} which we train on sets of signal models from \cite{Reis2020} and \cite{Reis2021}. It has been shown that global signal emulators such as \textsc{21cmGEM} \citep{Cohen2020} can be used for quick interpolation of the signal across the astrophysical parameter space \citep{Monsalve2019}. \textsc{globalemu} is a flexible framework that can easily learn different simulations of the global signal and has been shown to be faster and more accurate than the previous state of the art \citep{Cohen2020}. We provide more details on the accuracy of each trained instance of \textsc{globalemu} in \cref{sec:signal_modelling}.

We illustrate the presence of a sinusoidal systematic in the data and attempt to physically model the structure in a manner which is independent of the foreground model. We use two separate models each representing the introduction of a systematic at different points in the SARAS2 experiment. The motivation for each model is explained in \cref{sec:systematic modelling}.

The identification of the systematic is driven by the application of \textsc{maxsmooth} \citep{Bevins2020, Bevins2021} to model the foreground and smooth systematics in the data with a model that has constrained derivatives and resultant smooth properties based on Maximally Smooth Functions \citep[MSFs, ][]{Sathyanarayana2015, Sathyanarayana2017}. The motivation behind the use of \textsc{maxsmooth} is two-fold. Firstly, the SARAS2 antenna is designed and has been shown to have a maximally smooth reflection coefficient and efficiencies \citep{Singh2018a}.
Secondly, the dominant foregrounds in global 21-cm experiments from Galactic and extragalactic synchrotron and free-free emitting sources are expected to be smooth power laws \citep{Sathyanarayana2017, Bernardi2009, Nitu2021}.

In \cref{sec:data} we discuss briefly the SARAS2 experiment and the data that we are analysing. This is followed by a more detailed description of the modelling that we perform in \cref{sec:modelling}, a discussion about the sensitivity of the data to specific models in \cref{sec:sensitivity} and a summary of our results in \cref{sec:results}. We draw conclusions in \cref{sec:conclusions}.

\section{The SARAS2 data}
\label{sec:data}

One of the primary causes of systematics in global 21-cm experiments is chromaticity in the typically very wide beam pattern of the antenna. Further, sidelobes in the beam and complex reflection coefficients can also introduce frequency dependent structures in the data. The SARAS2 antenna is a short monopole designed to have an achromatic response. 

In principle, the foreground and systematics in the data from the SARAS2 experiment should both be smooth in nature and significant efforts were made to ensure that the efficiency and reflection coefficients in the data were smooth functions \citep{Singh2018a}. This has been explored further in \cite{Sathyanarayana2017} where it was shown that simulated observations with the SARAS2 antenna of the foregrounds \citep[produced with the Global Model for the Radio Sky Spectrum,][]{Sathyanarayana2016} in a global 21-cm experiment are smooth in structure to within a few mK. This is also expected generally, in the absence of ionospheric effects \citep{Shen2021}, for an achromatic beam like SARAS2.

SARAS2 is deployed in the remote radio quiet Timbaktu Collective in Southern India (lat = + 14.$^{\circ}$242328, long = 77.$^{\circ}$612606E). The antenna is comprised of a sphere mounted on top of an inverted cone resting on a circular aluminium disk. The components are smoothly joined tangentially and placed above the receiver electronics at the site. The electronics are battery powered and the site is flat and open. An optical fiber is used to connect the receiver to a signal processing unit situated 100 m away.

The beam pattern of the SARAS2 antenna is simulated, measured and shown to be frequency independent \citep{Singh2018a}. The pattern is omni-directional and constant in azimuth, with nulls at zenith and horizon, a peak at 30 degrees in elevation and a half-power beam width of 45 degrees in elevation. A 3D visualisation can be found in Fig. 8 of \cite{Sathyanarayana2017}. 

The antenna temperature, assuming the presence of a global 21-cm signal $T_{21}$, would correspond to
\begin{equation}
    T_\mathrm{A} = (T_{21} + T_\mathrm{gr} + T_\mathrm{fg})\eta_t,
\end{equation}
where $T_\mathrm{fg}$ accounts for the Galactic and extragalactic foregrounds and $\eta_t$ corresponds to the total efficiency of the SARAS2 antenna. $T_\mathrm{gr}$ refers to ground emission and for the analysis presented here we assume that the ground emission is smooth or equivalently that the ground under the antenna is homogeneous. As a result we can subsume the ground emission term into our smooth foreground model and treat the antenna temperature as being given by
\begin{equation}
    T_\mathrm{A} = (T_{21} + T_\mathrm{fg})\eta_t.
    \label{eq:components}
\end{equation}
Note that the assumption of a homogeneous ground under the antenna may not hold and that this may cause the introduction of non-smooth systematics into the data~(see \cref{sec:systematic modelling}). The sum $T_W = (T_{21} + T_\mathrm{fg})$ represents the beam-weighted sky power and $\eta_t$ is the product of the radiation and reflection efficiency \citep{Singh2018a}. $\eta_t$ therefore accounts for the loss due to an impedance mismatch between the antenna and the transmission line to the receiver as well as the frequency dependent coupling of the beam-weighted sky temperature to the antenna. Estimates of $\eta_t$ are made using the GMOSS simulations and measurements of the differential antenna temperature as the sky passes through the beam. The calibration and RFI rejection are detailed in section 6 of \cite{Singh2018a} and summarised in \cite{Singh2017}. We are assuming that the data has been calibrated to be in Kelvin units of antenna temperature and that there is no residual RFI.

The data can be seen in \cite{Singh2017, Singh2018} and we discuss the sensitivity of the data to the global 21-cm signal  in \cref{sec:sensitivity} after introducing the signal models in the following section.

\section{Modelling}
\label{sec:modelling}

The bayesian nested sampling tool \textsc{polychord} \citep{Handley2015a, Handley2015b} is used to fit two different systematic models and two different parameterisations of the global signal to the SARAS2 data. We model the foreground using the software \textsc{maxsmooth} \citep{Bevins2020, Bevins2021} and emulate physical models of the global signal with \textsc{globalemu} \citep[][]{Bevins2021b}.

In practice we model the foreground as $T_\mathrm{fg}^* = T_\mathrm{fg}\eta_t$. Throughout the rest of the paper we generally assume, unless otherwise stated, that when discussing the foreground we are including in that definition $\eta_t$ and any additive smooth systematics. We consider the addition of non-smooth systematics, $T_\mathrm{NS}$, into \cref{eq:components} in \cref{sec:systematic modelling}. The details of the different components of our model are given in the following sub sections. 

The Bayesian modelling techniques used here are increasingly common practice in 21-cm cosmology \citep[e.g.][]{Monsalve2019, 10.1093/mnras/staa487,Mondal2020, 10.1093/mnras/stab776, Chatterjee2021} and form the basis of the data analysis pipeline for REACH \citep[see][Sims et al. (in prep.)]{Anstey2021}. We briefly discuss the nested sampling algorithm and the reproducibility of our results in \cref{app:reproducability}.

\subsection{Noise Modelling}
\label{sec:noise_models}

\begin{table}
    \centering
    \begin{tabular}{|p{1.1cm}|p{1.3cm}|p{2.5cm}|p{1.5cm}|}
        \hline
        Noise Model & $\sigma$ & Prior & Prior Type \\
        \hline
        \hline
        Constant & $A_{\sigma}$ & $A_{\sigma}=10^{-3}-10^{-1}$ mK & Log Uniform\\
        \hline
        \multirow{2}{1.5cm}{Frequency Damped} & \multirow{2}{*}{$A_{\sigma}\bigg(\frac{\nu}{\nu_0}\bigg)^{-\beta_{\sigma}}$} & $A_{\sigma}=10^{-4}-10^{-1}$ mK & Log Uniform \\
        \cline{3-4}
        & & $\beta_{\sigma} = 0 - 5$ & Uniform \\
        \hline
         Relative Weights & $A_{\sigma}~W(\nu)$ & $A_{\sigma} = 10^{-2}-10^{-1}$ mK & Log Uniform\\
         \hline
    \end{tabular}
    \caption{The tested frequency dependent and independent standard deviation models for the assumed Gaussian noise in the SARAS2 data. In the frequency damped noise model $\nu_0$ is the central frequency in the band. The origin of the relative weights, $W(\nu)$, is discussed in \cref{sec:noise_models}.}
    \label{tab:noise_models}
\end{table}

For all of the fits performed in this paper, we assume that the noise in the data is Gaussian distributed and use a Gaussian log-likelihood function
\begin{equation}
    \log\mathcal{L} = \sum_i \bigg(-\frac{1}{2}\log(2\pi\sigma^2) -\frac{1}{2}\bigg(\frac{T_\mathrm{A}(\nu_i) - T_\mathrm{M}(\nu_i)}{\sigma}\bigg)^2\bigg),
    \label{eq:likelihood}
\end{equation}
where $T_\mathrm{M}$ stands for the sum of the model components described below. Our assumption is supported by assessment of the noise, which shows a Gaussian distribution, in data that has passed through the SARAS2 radiometer using a series of different terminations measured in the lab \citep{Singh2018a}. Support is also given by previous analysis of the data in which the residuals after foreground modelling with a high order polynomial have been shown to be Gaussian distributed \citep[see][ and \cref{app:residuals}]{Singh2017}.

Typically the noise is assumed to be frequency independent, however, in practice the noise is dependent on the system temperature which is dominated by the sky temperature and is a function of frequency. In this paper we consider three different approximations to the standard deviation, $\sigma$, of the assumed Gaussian noise each with a different frequency dependence. The first is a constant value of $\sigma$ and the second is given by a frequency damped amplitude. The latter comes from the naive expectation that $\sigma$ should be proportional to $T_W$ which means that the standard deviation should decrease with increasing frequency following the trend of the dominant foregrounds. Our third model uses the relative weights, $W(\nu)$, for the data which are dependent on the RFI excision, integration time and system temperature \citep[see Fig. 4 in ][]{Singh2018}. The noise models are summarised in \cref{tab:noise_models} and we discuss the results when fitting with the proposed models in \cref{sec:noise_results}. Previous analysis has indicated that the standard deviation on the noise is likely to be constant across the band in the calibrated and sky-averaged data \citep{Singh2017, Singh2018}.

A detailed study of likelihood and noise modelling in global 21-cm experiments is in preparation by \cite{Scheutwinkel2022}.

\subsection{Foreground Modelling}

Previously, the foreground in the SARAS2 data set has been modelled in combination with systematics using a high order polynomial \citep[$N = 4 - 8$, ][]{Singh2017, Singh2018}. However, while a polynomial will fit out both the foregrounds and smooth systematics, it could equally fit out part or all of any global signal and any non-smooth systematics in the data.

We model the foreground and smooth systematics using a variant of an MSF \citep{Sathyanarayana2015, Sathyanarayana2017} called a Partially Smooth Function \citep[PSFs,][]{Bevins2021}. An MSF has derivatives of order greater than or equal to two constrained so that the function does not have any inflection points or higher order non-smooth structure (i.e. the constrained derivatives do not cross zero in the band). PSFs are closely related to MSFs but more general in their definition and can have an arbitrary set of constrained derivatives\footnote{For clarity, the F in PSF and MSF stands for "function" and elaborate MSFs can be designed with exponential or trigonometric basis functions. However, the models are typically polynomial with a finite number of derivatives. A discussion of this is found in Section 2 of \cite{Bevins2021}.}.

The SARAS2 data has both a turning point and inflection point that can be attributed to the foreground multiplied by $\eta_t$ \citep[][]{Singh2017, Singh2018}. We, therefore, model the PSF foreground with derivatives of order $m \geqslant 3$ constrained according to
\begin{equation}
    \frac{d^m T_\mathrm{fg}^*}{d\nu^m} \leqslant 0~~\textnormal{or}~~\frac{d^m T_\mathrm{fg}^*}{d\nu^m} \geqslant 0,
\end{equation}
with the software \textsc{maxsmooth}. This prevents the introduction of high order non-smooth structure into the model but allows the foreground model to fit for a turning point (with $d T_\mathrm{fg}^*/d\nu = 0$ at some frequency, $\nu$, in the band) and inflection point (with $d^2 T_\mathrm{fg}^*/d\nu^2 = 0$).

We test the range of built-in \textsc{maxsmooth} foreground models \citep[see ][]{Bevins2021} and find that
\begin{equation}
    T_\mathrm{fg}^* = \sum^{N-1}_{k=0} a_k (\nu - \nu_0)^k,
    \label{eq:psf_fore}
\end{equation}
is the best fitting model with $N \geq 10$. Here $\nu_0$ is the central frequency across the bandwidth. Note that \textsc{maxsmooth} is not a Bayesian algorithm and the model parameters $a_k$ for the foreground are not fitted by \textsc{polychord}. Instead, we wrap \textsc{maxsmooth} inside the call to \textsc{polychord} and at each sample point \textsc{maxsmooth} fits the foreground parameters, $a_k$, to $T_\mathrm{A} - T_{21}\eta_t - T_\mathrm{NS}$.

\begin{figure}
    \centering
    \includegraphics{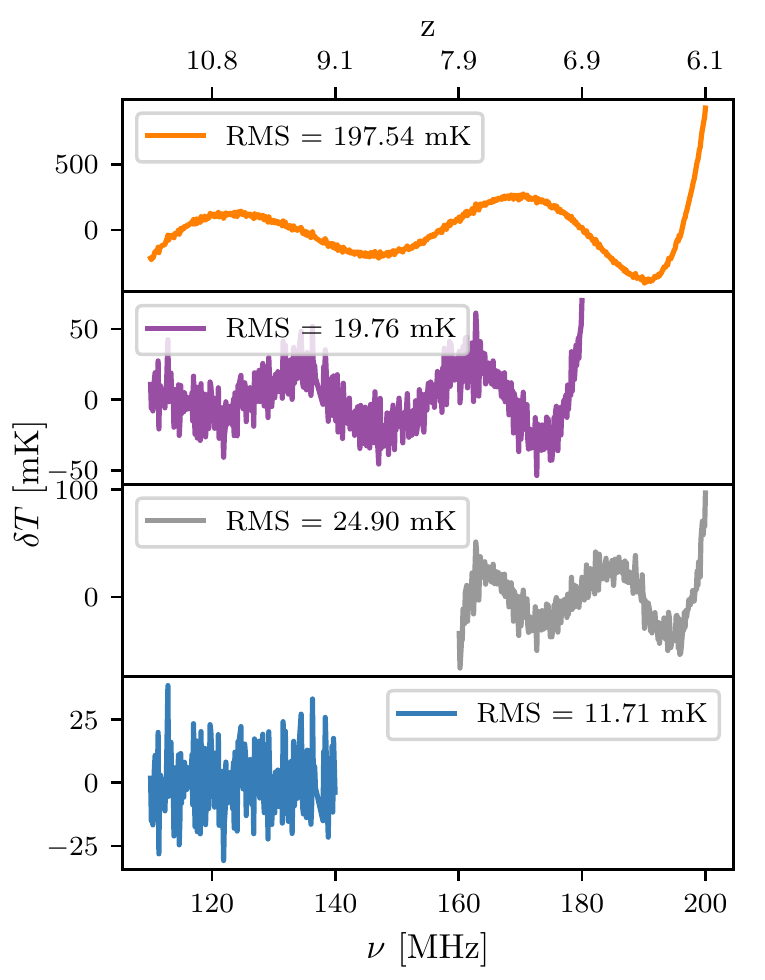}
    \caption{A comparison of the residuals when fitting the SARAS2 data with a $10^{th}$ order Partially Smooth Function with derivatives of order $m \geqslant 3$ constrained across the full SARAS2 band (orange) and a set of reduced bandwidths. We achieve a significantly lower RMS in the reduced bandwidths potentially improving the signal to noise ratio in the data. This indicates either a poor foreground fit across the full bandwidth, which can introduce non-smooth structure, or the presence of multiple non-smooth systematics dominant at different frequencies. We proceed to perform our analysis in the reduced bandwidth 110-180 MHz based on the results presented in previous work \protect\citep{Singh2017,Singh2018a} and in order to retain as much data as possible.}
    \label{fig:foreground_res}
\end{figure}

\cref{fig:foreground_res} shows the resultant residuals, in orange, when fitting the data with the PSF model across the whole SARAS2 bandwidth. The residuals are large in magnitude and show a sinusoidal structure which may be the result of systematics in the data and/or of inaccurate foreground modelling. 

In the previous analysis of the SARAS2 data, in which parameter constraints were determined from a discrete set of signal models, the bandwidth used was optimised on a per-signal basis \citep{Singh2017, Singh2018}. In that work it was frequently found that the bandwidth $110-180$ MHz is the optimum to minimise the signal-to-noise ratio for the 264 tested signal models

In this work, the turning point and inflection point may be significantly distorting the foreground model leading to the introduction of spurious non-smooth structure in the residuals\footnote{This is unlikely and it can actually be shown that Partially Smooth Functions can be effectively used to recover the noise in data sets that feature inflection points (see `Turning Points and Inflection Points' in \url{https://maxsmooth.readthedocs.io/en/latest/maxsmooth.html}).}, despite allowing for their presence in the modelling, and so we attempt to fit with a PSF in the reduced range $\nu = 110 - 180$ MHz effectively removing the turning point. We achieve significantly smaller residuals when fitting in this band as shown in purple in \cref{fig:foreground_res}. The difference in magnitude may suggest a better quality foreground fit in the band $110 - 180$ MHz but it may also indicate the presence of multiple non-smooth systematics in the data each of which may dominate to a different degree at different frequencies. 

A further reduction in the upper bound on the frequency range leads to a further reduction in the RMS. However, we still see the same sinusoidal structure at low frequencies. In practice we could reduce the band to 110 - 140 MHz, removing both the turning point and inflection point, which approximates to one full cycle of the sinusoidal structure in both sets of residuals in the two upper panels of \cref{fig:foreground_res} and we would see a significantly lower RMS ($\approx 12$ mK when fitting with the proposed PSF model as can be seen in the bottom panel of \cref{fig:foreground_res}). This is because the systematic structure in the data in this frequency range covers one period of oscillation and is therefore smooth to a sufficient level that it is removed by the foreground model. Further, if we remove the inflection point in the data and fit in the range $160 - 200$ MHz we see a sinusoidal structure that is partially consistent with the purple residuals in \cref{fig:foreground_res}.

Throughout the rest of the paper, in accordance with the previous SARAS2 analysis, to simplify the modelling and to keep as much data as possible, we use the reduced bandwidth $110 - 180$ MHz. The modelling of the non-smooth systematic structure is considered in the following section.

We note that contributions from foreground polarization can generally be expected in the data. Modelling of the effects of polarization is non-trivial, however, and the intensity of corresponding contributions is dependent on a number of factors \citep{Spinelli_polarization_2018}. In \cite{Spinelli_polarization_2019} the authors show that contributions from polarisation have significant non-smooth structure and we would expect that, if this was present and dominant in our data, it would be obvious after modelling the foreground with a PSF. However, this is not the case in our residuals. Further, to a first approximation we expect that the polarized signal will be proportional to $1/\nu^2$ \citep{Spinelli_polarization_2018} and larger at lower frequencies following the opposite trend to our residuals. Therefore, any contribution from foreground polarization in our data can be considered sub-dominant and subsumed in our noise modelling.

In comparison to the predicted maximum depth of the global 21-cm signal \citep[$\approx 165$ mK for standard astrophysical signals in the band $110 - 180$ MHz,][]{Reis2021}, the residual RMS, $19.8$ mK, appears small in magnitude. However, as highlighted above and in \cite{Singh2017} and \cite{Singh2018}, any signal in the data will be suppressed by the total efficiency of the antenna. This is discussed further in \cref{sec:signal_modelling}. 

\subsection{Systematic Modelling}
\label{sec:systematic modelling}

\begin{figure}
    \centering
    \includegraphics[width=\linewidth]{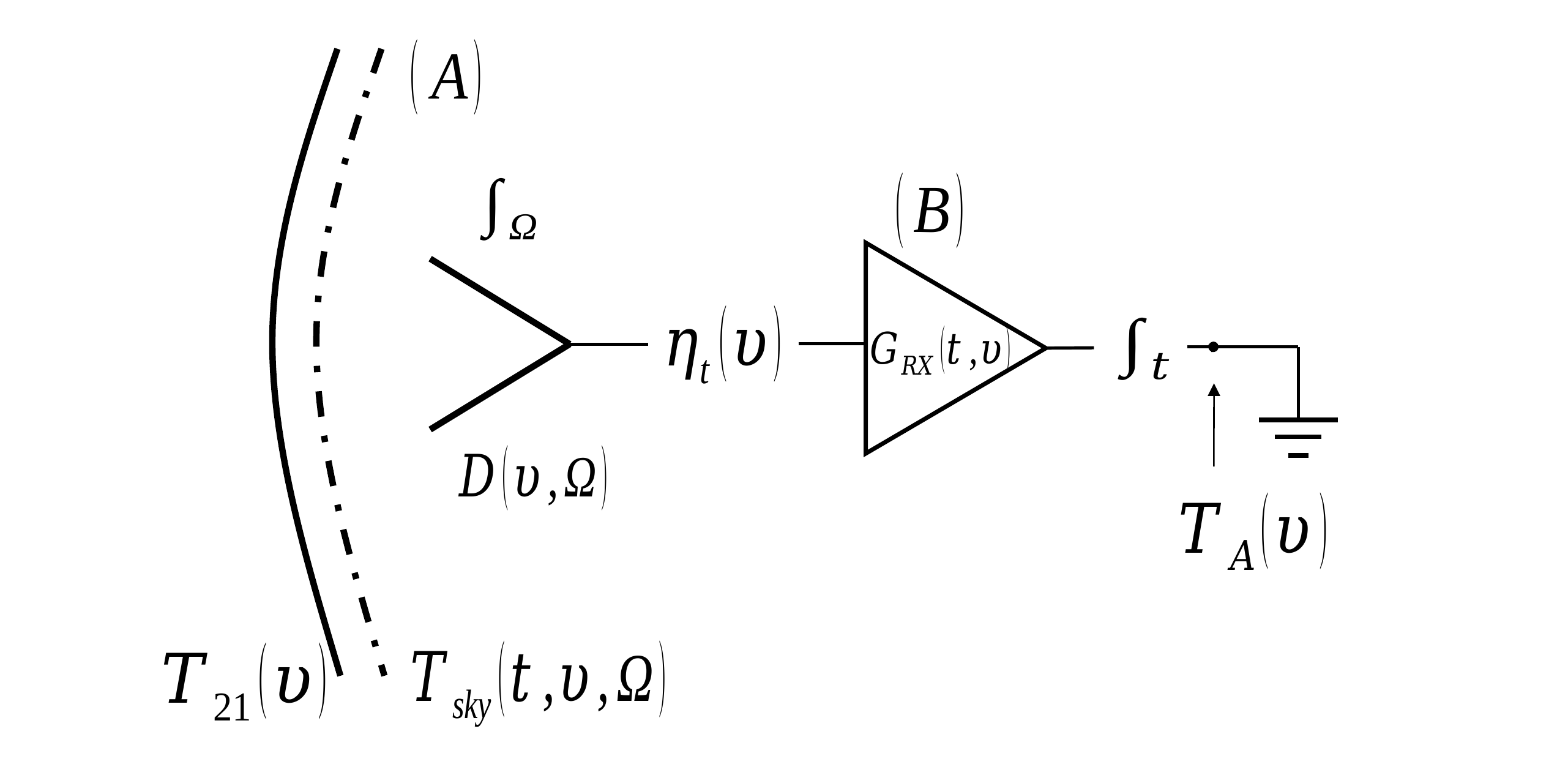}
    \caption{A simplified schematic of a global 21-cm experiment. The figure illustrates the two positions at which systematic power could be added into $T_\mathrm{A}$; (A) prior to the antenna and (B) after the antenna in the receiver electronics denoted by the time dependent gain, $G_\mathrm{RX}(t, \nu)$. Note when modelling the systematic in the SARAS2 data we are considering these scenarios independently. The former systematic arrives with the signal from the sky, $T_\mathrm{sky}(t, \nu, \Omega)$, and global signal, $T_{21}(\nu)$ if present, and passes through the antenna where it is weighted by the beam directivity, $D(\nu, \Omega)$. The angular dependence is integrated out and the systematic is then multiplied by the total efficiency of the antenna, $\eta_t(\nu)$, giving $T_\mathrm{NS,2}$ as detailed in \cref{sec:systematic modelling}. The systematic introduced in the electronics corresponds to $T_\mathrm{NS,1}$ in \cref{sec:systematic modelling}. After passing through the electronics the time dependence is also integrated out of the data leaving $T_A$. Figure modified from Fig.2 of \protect\cite{Cumner2021}.}
    \label{fig:radiometer}
\end{figure}

Previous analysis has shown the potential presence of sinusoidal and damped sinusoidal systematics in data from several global 21-cm experiments using a variety of different radiometers and analysis techniques. One such example, previously mentioned in the introduction, is the EDGES data. \cite{Hills2018} used a foreground model with spectral index characterised by a 6 term unconstrained polynomial to identify a 60 mK sinusoidal structure in the EDGES data. A subsequent re-analysis of the data in \cite{Singh2019} and \cite{Bevins2021}, using MSFs to model the foregrounds, identified a similar sinusoidal structure. In \cite{Bevins2021} the authors also re-analysed data from LEDA and identified the presence of a damped sinusoidal systematic structure using MSFs. This conclusion is supported by previous investigation of the LEDA data in \cite{Price2018} where the authors used a log-polynomial foreground model.

Currently, systematics and the ambiguity of their causes pose a limiting factor in the detection and/or confidence of a detection in global 21-cm experiments. We attempt to physically motivate our model for the systematic structure in the SARAS2 data and, as a consequence of detailed systematic modelling, derive astrophysical parameter constraints on the global signal.

Specifically, we model the damped sinusoidal structure seen in the residuals after foreground subtraction with two different physically motivated systematic models representing the introduction of power at different points, (A) and (B), in the SARAS2 system as illustrated in \cref{fig:radiometer}. Any non-smooth systematic model can be included in \cref{eq:components} as an additive term
\begin{equation}
    T_\mathrm{A} = (T_{21} + T_\mathrm{fg})\eta_t + T_\mathrm{NS}.
    \label{eq:components_with_sys}
\end{equation}
The sinusoidal structure may be introduced by a poor estimate of $\eta_t$. However, when fitting the total efficiency with a best fitting MSF of the form given in \cref{eq:psf_fore} we find no sinusoidal structure in the residuals.
We also reiterate, for completeness, that the systematic structure may be being introduced by poor foreground modelling.

However, the first systematic model, which we refer to as the damped systematic model, corresponds to power introduced after the antenna and in the electronics \citep[e.g. see Fig. 2 in ][]{Singh2018a}
\begin{equation}
    T_\mathrm{NS, 1}(\nu) = \bigg(\frac{\nu}{\nu_0}\bigg)^{\alpha_\mathrm{sys}} A \sin\bigg(\frac{2\pi\nu}{P} + \phi\bigg),
    \label{eq:damped_sys}
\end{equation}
where $\alpha_\mathrm{sys}$ is a damping power, $A$ is the amplitude of the systematic, $P$ is the period, $\phi$ is the phase and $\nu_0$ is the central frequency in the band. We note that the prior range of $\alpha_\mathrm{sys}$ ranges from $0 - 10$ (see \cref{tab:priors} for all of the systematic prior ranges) and as a result the model can also account for sinusoidal systematics. The fitted parameters, $A$, $\alpha_\mathrm{sys}$, $P$ and $\phi$ are constant across the band and the frequency dependence comes from the damping factor, $(\nu/\nu_0)^{\alpha_\mathrm{sys}}$.

The second model, the efficiency systematic, is given by
\begin{equation}
    T_\mathrm{NS, 2}(\nu) = \eta_t \bigg(\frac{\nu}{\nu_0}\bigg)^{\alpha_\mathrm{sys}} A \sin\bigg(\frac{2\pi\nu}{P} + \phi\bigg).
    \label{eq:eff_sys}
\end{equation}
In this case, the systematic structure models power introduced prior to the antenna and mediated by the total efficiency which provides some damping. Such a systematic could be explained by activity in the ionosphere over the observing period, RFI or a previously unidentified non-smooth component of the foreground.

The non-smooth structure in the data could also, as highlighted previously, be caused by emission from the ground if the assumption that this is smooth does not hold. Specifically, structure or layering in the ground at depths larger than the wavelengths of operation and below the penetration depth computed for the soil properties could introduce systematic structure. A discontinuity in the soil below the antenna such as that between the loose top soil, caused by erosion over time, and the rock of the Deccan Plateau or from a water table could cause a damped sinusoidal structure to propagate through to the receiver noise, total efficiency and antenna temperature.

Another possible origin of a damped sinusoidal systematic could be a small clump of foliage or a root system, without significant foliage above ground, close to the deployment sight. In principle both this and the above ground emission are potential causes of systematics that will be mitigated in the latest iteration of the SARAS experiment, SARAS3, which has been deployed on a lake \citep{Girish2020, SARAS2021, Raghunathan2021}.

It is well-known that the directivity of vertical monopoles ~\citep{5492282} naturally experiences a sinusoidal like frequency response. Such behaviour would correspond to $T_\mathrm{NS, 2}$. We note, however, that the period of the sinusoidal-like structure visible in the data is faster than what would be expected from a monopole. The scale of the circular element of the antenna, 43.5 cm in radius, was chosen such that any reflections from the edges of the disks would have a period of $\approx 350$ MHz and the observing band falls within the first resonance at $260$ MHz \citep{Singh2017, Singh2018a}. Any reflections in the beam pattern from the edges of the disk would thus be smooth across the reduced SARAS2 band and effectively subsumed by our smooth foreground model detailed in the previous subsection.

If we consider the systematic to be real and not a spurious signal introduced by an inaccurate foreground model, then our analysis could help to identify a cause because we have two distinct models representing systematics introduced at different points in the experiment. We note that there may be degeneracy between the two different models but also that the efficiency is an inherent characteristic of the experiment and unlikely to mimic generic systematic properties.

\begin{table}
    \centering
    \begin{tabular}{|P{1.2cm}|P{1.3cm}|P{3cm}|P{1.4cm}|}
        \hline
         & Parameter & Prior & Prior Type\\
        \hline
        \hline
        \multirow{4}{*}{Systematic}& $\alpha_\mathrm{sys}$ & $0 - 10$ & \multirow{4}{*}{Uniform}\\
        & $A$ & $0 - 1$ K  & \\
        & $P$ & $10 - 70$ MHz &  \\
        & $\phi$ & $0 - 2\pi$ rad & \\
        \hline
        \hline
        \multirow{8}{*}{Signal}& $\tau$  & $0.026 - 0.1$ (STA) / $0.035 - 0.077$ (ERB) & \multirow{4}{*}{Uniform}\\
        & $\alpha$ & 1.3 (STA only) & \\
        & $E_\mathrm{min}$ & $0.1 - 3$ keV (STA only) &\\
        & $R_\mathrm{mfp}$ & 30 (STA) / 40 (ERB) Mpc &\\
        \cline{2-4}
        & $f_*$ & $0.001 - 0.5$ &  \multirow{4}{*}{Log-Uniform}\\
        & $V_c$ & $4.2 - 100$ km/s & \\
        & $f_X$ & $0.0001 - 1000$ & \\ 
        &$f_\mathrm{radio}$ & $1 - 99500$ (ERB only) & \\
        \hline
    \end{tabular}
    \caption{The prior ranges and prior types used for the systematic and signal parameters fitted by \textsc{polychord}. Note that for the signal parameters the prior ranges are defined by the training data. For the excess radio background (ERB) signals $R_\mathrm{rmfp}$ is fixed at 40 Mpc and the X-ray SED is representative of that from X-ray binaries. See \cref{sec:signal_modelling} for more details on each model component, the training data and the difference between the standard astrophysical (STA) and ERB models.}
    \label{tab:priors}
\end{table}

\subsection{Signal Modelling}
\label{sec:signal_modelling}

The signal emulator \textsc{globalemu} provides a framework to train neural networks on different sets of global 21-cm signals. We can therefore use the latest simulations with the most up-to-date understanding of the signal and we can analyse parameter constraints on different astrophysical models. 

\begin{figure*}
    \centering
    \includegraphics{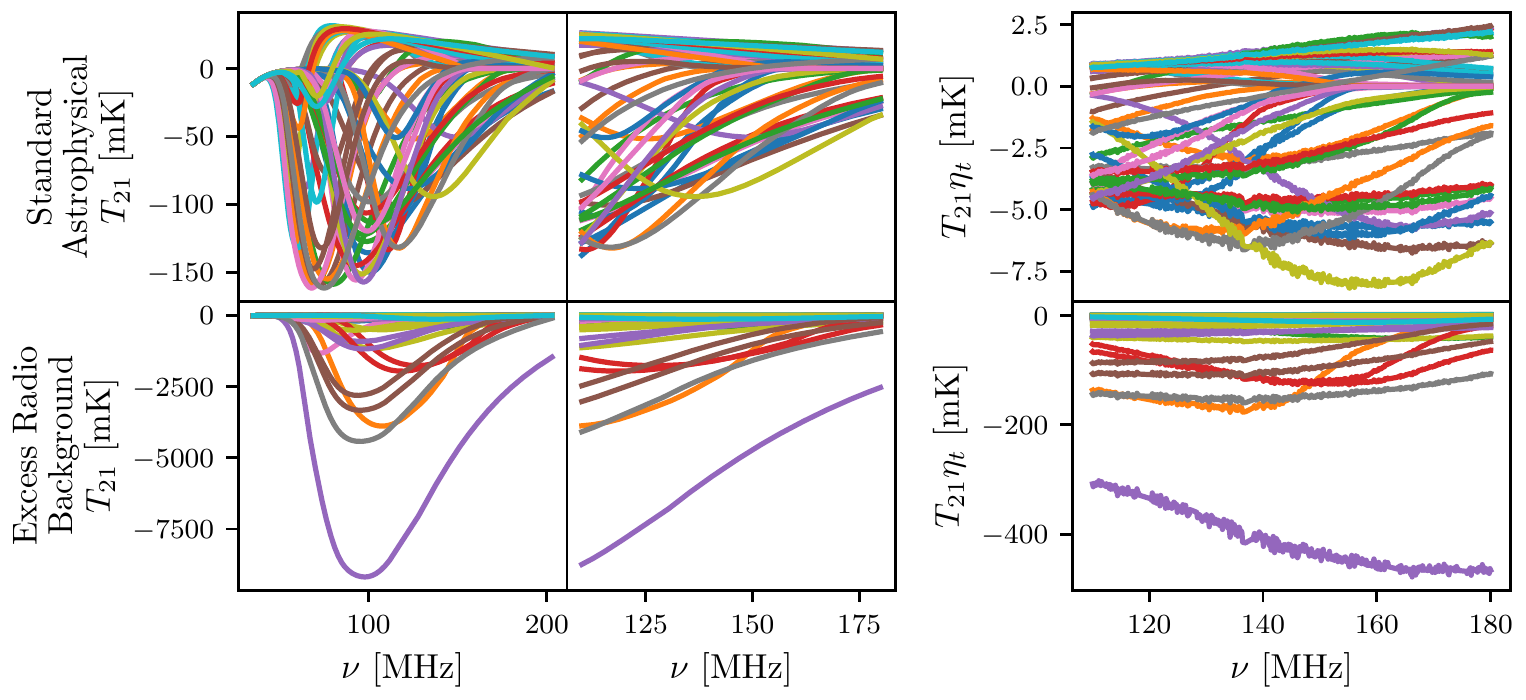}
    \caption{\textbf{Top Row, left to right:} A set of 50 example signals that include Lyman-$\alpha$ heating and CMB heating, the equivalent signals in the band $110 - 180$ MHz and the signals multiplied by the total efficiency of the SARAS2 antenna. \textbf{Bottom Row, left to right:} 50 example signals with different levels of excess radio background above the CMB, the same signals in the reduced SARAS2 band and finally the signals multiplied by the total efficiency of the SARAS2 antenna.}
    \label{fig:signals}
\end{figure*}

The global 21-cm signal is determined by the contrast between the spin temperature of neutral hydrogen, $T_s$, and the radio background, $T_r$, as a function of redshift
\begin{equation}
    T_{21} = \frac{T_s -T_r}{1+z}(1 - e^{-\tau_{21}}),
    \label{eq:deltaT}
\end{equation}
where $\tau_{21}$ is the 21-cm optical depth of the IGM \citep{Mesinger2019}. Calculated realisations of the global 21-cm signal are achieved by using hybrid techniques and either averaging over large modelled cosmological volumes \citep[2D simulations e.g.][]{Mesinger2011} or directly approximating that average over redshift \citep[1D simulations e.g.][]{Mirocha2014}. In this analysis we use the simulations detailed in \cite{Visbal2012, Fialkov2014, Cohen2017, Reis2020, Reis2021} and specifically the sets of global 21-cm signals presented in \cite{Reis2020} and \cite{Reis2021}. For each set of signals, exotic models with an excess radio background and standard astrophysics respectively, we train neural network emulators with \textsc{globalemu}. More details about the two sets of signal models we have used are given in the following sections with examples shown in \cref{fig:signals}.

\subsubsection{Standard astrophysical signals}

In this paper, we use the recent simulations of the global signal with Lyman-$\alpha$ heating, CMB heating and multiple scattering presented in \cite{Reis2021}. We refer to these standard astrophysical models in the text as the STA (standard astrophysical) signals. As stars form in the early universe and Lyman-$\alpha$ coupling between the spin temperature, $T_s$, of the neutral hydrogen and kinetic temperature of the gas, $T_k$, begins to influence the structure of global signal, Lyman-$\alpha$ and CMB heating begin to counteract the cooling \citep[][]{Chuzhoy2007, Venumadhav2018, Mittal2020, Reis2021}. These heating mechanisms begin before the onset of X-ray heating and they lead to a reduction in the theoretical maximum potential depth of absorption from $\approx250$ mK \citep{Cohen2020} to $\approx 165$ mK \citep{Reis2021}. Multiple scattering influences the efficiency of Lyman-$\alpha$ coupling, has a weak affect on the maximum depth of the signal and primarily influences the structure of the power spectrum at $z > 20$ \citep{Reis2021} outside the SARAS2 band.

We train \textsc{globalemu} on a set of 5,137 realisations of the standard astrophysical signals and test the quality of emulation with a set of 570 models. Each model spans the redshift range $z = 6 - 39$ and the redshift spacing is given by $\delta z = 0.1$. The structure of the signals is determined by a set of seven astrophysical parameters \citep[further details can be found in ][]{Cohen2020, Reis2021, Bevins2021b};

\begin{itemize}
    \item The star formation efficiency, $f_*$: Determines the fraction of gas in dark matter halos that is converted into stars. The value of $f_*$ drives the Lyman-$\alpha$ flux and influences the onset of X-ray heating which both determine the depth of the global signal absorption trough and the ionizing efficiency of sources.
    \item The minimal circular velocity, $V_c$: The threshold virial circular velocity is proportional to the cube root of the minimum halo mass for star formation. Its value is determined by different
    cooling channels (molecular and atomic hydrogen) and star-formation-suppressing feedback mechanisms. It influences the timing of Lyman-$\alpha$ coupling, total X-ray luminosity of halos and reionization.
    \item The X-ray efficiency, $f_X$: The value of $f_X$ affects the X-ray luminosity per star formation rate, which in turn influences the depth of the absorption trough and amplitude of any emission above the radio background during reionization. Further, $f_X$ has a minor effect on ionization at recent times.
    \item The slope of the X-ray spectral energy distribution~(SED), $\alpha$: The dependence of the structure of the global signal on $\alpha$ is expected to be very weak \citep{Monsalve2019} and its value plays a small role at low redshifts.
    \item The low energy cutoff of the X-ray SED, $E_\mathrm{min}$: This parameter regulates the fortness of the X-ray SED and thus has some influence on the efficiency of X-ray heating at low redshifts covered by the SARAS2 band. While it is not expected to affect the structure of the signal significantly, in some cases, it can have more of an impact than $\alpha$ or $R_\mathrm{mfp}$.
    \item CMB optical depth, $\tau$: The optical depth is directly related to the ionizing efficiency of sources, $\zeta$ and its value strongly influences the redshift of reionization. The value of $\tau$ has been determined by \cite{Planck2018} to be $0.055 \pm 0.007$ and \cref{tab:priors} shows that this range is explored completely by the prior for the STA models.
    \item The mean free path of ionizing photons, $R_\mathrm{mfp}$: The value of $R_\mathrm{mfp}$ affects the rate of ionization of the neutral hydrogen gas corresponding to the gradient of the signal at low redshifts. The effects of varying $R_\mathrm{mfp}$ are also not expected to influence the structure of the global signal significantly \citep[see e.g. ][]{Monsalve2019}.
\end{itemize}

The primary reason $E_\mathrm{min}$, $\alpha$ and $R_\mathrm{mfp}$, are explored in the simulations is because the simulations are also used to determine models of the power spectrum on which they have a greater influence. We train \textsc{globalemu} using all seven astrophysical parameters and subsequently perform fits with fixed values of $R_\mathrm{mfp} = 30$ Mpc and $\alpha = 1.3$ \citep[as was done with the EDGES High Band data in ][ using the contemporary standard astrophysical models and signal emulator]{Monsalve2019}. We repeat the analysis in \cref{app:lyman_res_vra} allowing \textsc{polychord} to fit for all seven parameters however we do not discuss these results in the main text as we find the effect of including the additional parameters is minimal. The ranges, equivalent to the priors, of all of the parameters sampled in the training and test data sets are given in \cref{tab:priors}.

We assess the accuracy of the emulator across the band $110 - 180$ MHz (or equivalently $z \approx 7 - 12$). We use a pragmatic target accuracy when emulating the signals of on average approximately 10\% of the expected noise from a global 21-cm experiment \citep{Bevins2021b}. This varies based on the experiment with a value of 2.5 mK for REACH, 2 mK for EDGES \citep{Bowman2018} and approximately 1 mK for SARAS2. Therefore, when emulating the STA models we assume a target accuracy of between 1 and 2.5 mK and use the RMSE metric given by equation (7) in \cite{Bevins2021b}. The mean, 95$\textsuperscript{th}$ percentile and the worst RMSE values from the test data set of 570 models, with sampling resolution equivalent to 0.122 MHz, are 0.8, 1.9 and 6.8 mK respectively. We find that only 29 models have an RMSE larger than $\approx 1.9$ mK indicating a high degree of accuracy. We use a fully connected network with 4 hidden layers of 16 nodes each to emulate the STA models.

In practice a low RMSE in temperature does not necessarily correspond to an accurate recovery of the astrophysical parameters. For the analysis performed in this work this is not a significant issue because the reported constraints are weak. We leave a detailed exploration of emulator accuracy and its affects on parameter recovery for future work.

\subsubsection{Excess radio background signals}

Typically, $T_r$ in \cref{eq:deltaT} is assumed to be equal to the CMB temperature. However, one of the possible explanations for the larger than expected absorption feature reported by EDGES \citep{Bowman2018} is an excess radio background above the CMB. In addition, a population of radio sources at high redshifts could naturally contribute to $T_r$ \citep[e.g][]{Feng2018}. While there is some evidence for a larger than expected radio background from ARCADE2 \citep{fixsen_arcade_2011} and LWA \citep{dowell_radio_2018}, there remain some concerns about the Galactic modelling in these works \citep{Subrahmanyan2013}.

\cite{Reis2020} investigated the introduction of an excess radio background from high redshift radio galaxies and we use the models presented there in this work in an attempt to constrain the parameter $f_\mathrm{radio}$ with the SARAS2 data. $f_\mathrm{radio}$ denotes the radio production efficiency of early galaxies, a value of one corresponds to the present day and the range of $f_\mathrm{radio}$ in our training data set is given in \cref{tab:priors}. We refer to these exotic astrophysical models throughout the rest of the paper as the ERB (excess radio background) models.

The simulations use a similar parameter description of the global signal presented in the previous subsection with the additional parameter $f_\mathrm{radio}$. The value of $R_\mathrm{mfp}$ is fixed at $40$~Mpc when running the simulations. The X-ray SED is assumed to be from X-ray binaries \citep{fialkov_observable_2014} and the simulations are consequently independent of $\alpha$ and $E_\mathrm{min}$. The models also include the effects of Lyman-$\alpha$ heating, CMB heating and multiple scattering. 

The data set contains 4,311 training models and 479 test models. We train \textsc{globalemu} with the five parameters as inputs; $f_*$, $V_c$, $f_X$, $\tau$ and $f_\mathrm{radio}$. The mean, 95th percentile and the worst RMSE values are 7.3, 27.3 and 125.9 mK respectively. We use a network with 4 hidden layers each with 16 nodes as was done with the STA models and note that with this training data set the accuracy does not significantly improve if we increase the size of the network. However, the mean accuracy of emulation, $7.3$ mK, is within an order of magnitude of our target accuracy of $1 - 2.5$ mK. The magnitudes of the signals, after multiplication by $\eta_t$, with the largest RMSE values are significant in comparison to the expected noise and as fractional accuracies the worst and 95th percentile results are reasonable.

\section{Sensitivity of the SARAS2 Data to Astrophysical Parameters}
\label{sec:sensitivity}

When considering the sensitivity of the SARAS2 data to the global 21-cm signal it is important to consider that any signal in the data will be multiplied by the total efficiency of the antenna. As has previously been discussed and shown in \cref{fig:signals}, this significantly reduces the magnitude of the signals in the data and given the expected noise\footnote{Specifically, this is the expected noise in the instrument plane without correcting for the total efficiency of the antenna.} of $11$ mK results in a low signal to noise ratio. In turn, this would make any signal in the data hard to recover.

Further, the observations cover a bandwidth expected to include the EoR which has implications for the types of signals that we expect to be able to constrain. The EoR window is sensitive to the value of $\tau$ which affects the redshift of reionization and as a result the maximum amplitude of the global signal. For example, a high value of $\tau$ can lead to an earlier reionization. In the absence of efficient X-ray heating and presence of a low Lyman-$\alpha$ flux the effect of $\tau$ is less significant in the SARAS2 bandwidth.

Similarly, the value of $E_\mathrm{min}$ has a more significant effect on the structure of the global signal in the SARAS2 band if $f_X$ is high. In this case the value of $E_\mathrm{min}$ affects the depth of the signal and the efficiency of X-ray heating. If we maintain the Lyman-$\alpha$ flux, i.e. onset of Lyman-$\alpha$ coupling, and increase the value of $E_\mathrm{min}$ we reduce the efficiency of X-ray heating, move the minimum of the signal to lower redshifts and can create a deep global signal inside the SARAS2 band. A similar effect can occur if we decrease the value of $f_X$ from high to low and this is more prominent than that introduced by variation in $E_\mathrm{min}$. The analysis should therefore be sensitive to models with low values of $f_X$ and high values of $E_\mathrm{min}$.

In addition, if $f_X$ is high, and to a lesser extent if $E_\mathrm{min}$ is low, then we can expect that there will be significant excess radio background which will produce a prominent and deep signal in the SARAS2 data even after multiplication by the total efficiency of the antenna.

\begin{figure*}
    \centering
    \includegraphics{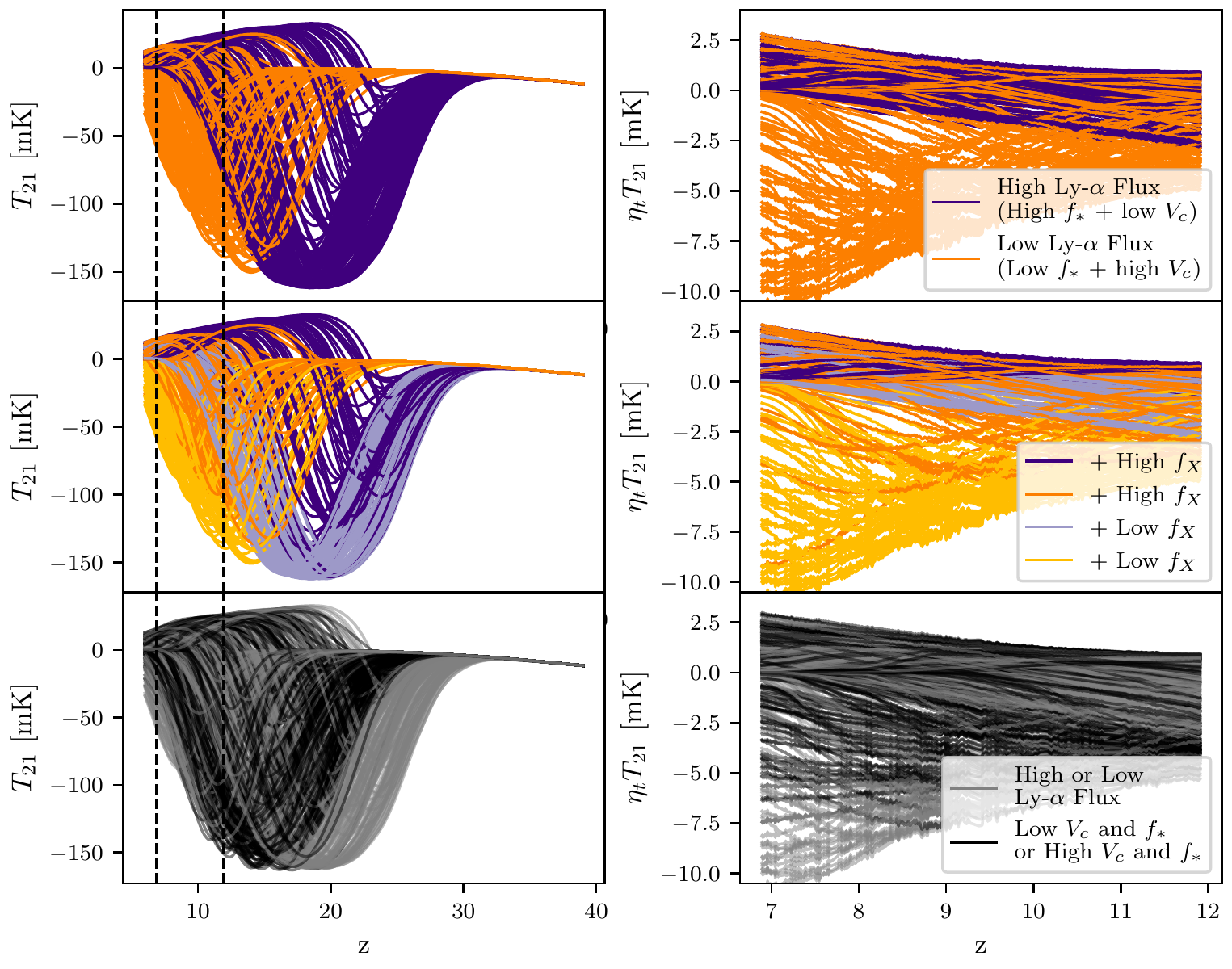}
    \caption{The figure shows 243 STA signals from the training data coloured based on the values of $f_*$, $V_c$ and $f_X$. The left column shows the signals as they are in the training and testing data and the right panel shows the signals in the SARAS2 band, indicated on the left by the dashed lines, after multiplication by the total efficiency of the antenna. In the top panels we show the set of signals split based on the type of Lyman-$\alpha$ flux that they correspond to. A high Lyman-$\alpha$ flux is produced by a high $f_*$ and low $V_c$ where we judge "low" and "high" with respect to the middle of the log-prior range for each parameter. In the middle panels we further split the models based on their values of $f_X$. From the graph we can see that, after multiplying by the total efficiency of the SARAS2 antenna, the signals with a low Lyman-$\alpha$ flux and low value of $f_X$ have the largest absolute magnitudes and so our analysis should be most sensitive to these models. The bottom panel shows the previously classified 243 STA signals in grey in comparison to a further 257 signals in black that do not meet our crude classification of "low" and "high" Lyman-$\alpha$ fluxes and with neither high $f_*$ and low $V_c$ or low $f_*$ and high $V_c$. These models have prominent structures in the SARAS2 band indicating a more general sensitivity to the values of $f_*$ and $V_c$.}
    \label{fig:sensitivity}
\end{figure*}

$V_c$ and $f_*$ determine the strength of the Lyman-$\alpha$ flux, consequent onset of Lyman-$\alpha$ coupling and position of the signal minimum. High values of $f_*$ correspond to a high fraction of the gas in dark matter halos being converted into stars which leads to a high Lyman-$\alpha$ flux. The top panels of \cref{fig:sensitivity} show that the strongest signals in the SARAS2 band after multiplication by the total efficiency are those with low Lyman-$\alpha$ fluxes (high $V_c$ and low $f_*$) whereas the weakest signals are those with high Lyman-$\alpha$ fluxes (low $V_c$ and high $f_*$). Here we have defined "high" and "low" values of the parameters with respect to the middle of the log-prior ranges. Our analysis should therefore be more sensitive to models with low Lyman-$\alpha$ fluxes.

Further dividing the two classes of signals in \cref{fig:sensitivity} based on their value of $f_X$ we can see that we should also expect our analysis to be sensitive to low values of $f_X$, as expected from the discussion above.

We should also consider our sensitivity to models that have both high or both low values of $V_c$ and $f_*$ in combinations that do not meet the crude criteria for high and low Lyman-$\alpha$ fluxes defined above. These models are shown in black in the bottom panel of \cref{fig:sensitivity} against the backdrop of previously "classified" models in grey. From the figure we can see that these models typically also have low Lyman-$\alpha$ fluxes and have minima at more recent redshifts particularly in comparison to the "high" Lyman-$\alpha$ flux signals discussed previously. These signals therefore have dominant structures and relatively large magnitudes in the SARAS2 band and indicate a more general sensitivity to the values of $f_*$ and $V_c$.

Finally, the data is expected to be sensitive to ERB signals with very high values of $f_\mathrm{radio}$ as these signals have deep absorption troughs of a few hundred mK even after multiplication by the total efficiency of the antenna. This is particularly true when there is also a high Lyman-$\alpha$ flux and low X-ray efficiency which results in strong variation of the signal within the SARAS2 band.

\section{Results}
\label{sec:results}

\begin{figure*}
    \centering
    \includegraphics{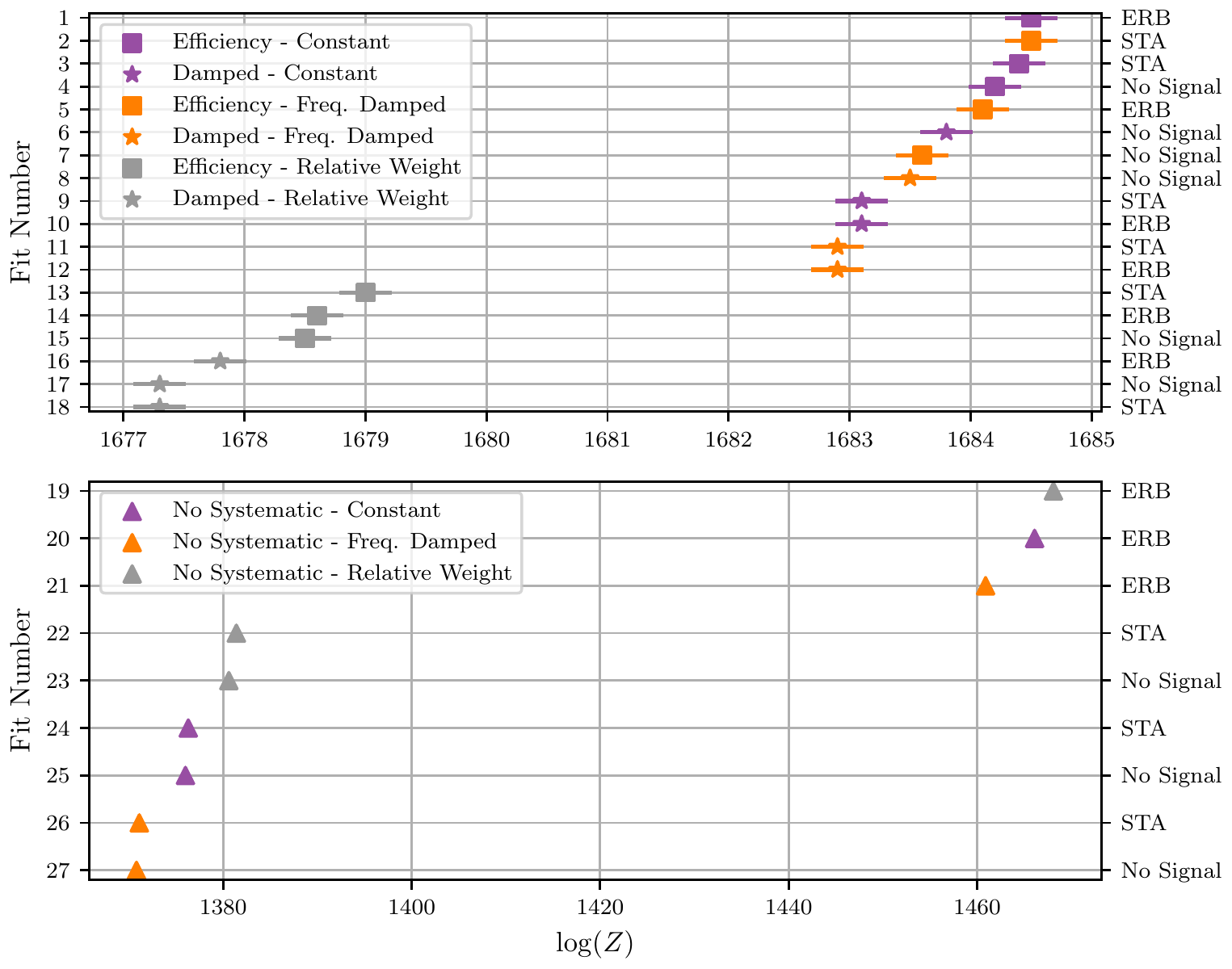}
    \caption{A graphical summary of the different fits performed in this analysis which demonstrates that the highest evidence fits from all of those performed are fits of equal quality. The signal type can be found on the right-hand y-axis, the fit number on the left-hand axis and the noise and systematic types in the legends. The fit numbers are used for referencing through out the text. The top panel shows fits that include a systematic model and the bottom panel shows fits without a systematic model. We divide the two classes of fits into separate graphs because those with a systematic have a much higher evidence, $\Delta \log(Z) \approx 200-300$, than those without. Note that a difference of $\Delta \log(Z) \approx 10$ corresponds to a difference of $\exp(10)\approx 22000$ or betting odds of 22000:1 in favour of the higher evidence model.}
    \label{fig:model_fits_graph}
\end{figure*}

\Cref{fig:model_fits_graph} summarises the different combinations of signal, systematic and noise models that were fitted to the SARAS2 data. For completeness, we report fits without signals and/or systematics. The four highest evidence fits are approximately equivalent (within errorbars) and the relative weights based noise model is comparatively poor compared to the two alternatives considered when we include a systematic model in the fit. These points are further discussed in the following sections.

The evidence, $Z$, is a marginal likelihood integrated over all of the fitted parameters. It quantifies the probability that the data is described by the chosen model components and is the normalising factor in Bayes theorem. A higher $\log(Z)$ indicates a preference for that model or hypothesis as a description for the data over alternatives. It is often used to determine the presence or absence of signals in data sets by comparing its value for fits with and without the relevant model components. An example of this, beyond the results presented in this paper, can be found in \cite{Bevins2021} in which the authors fit the EDGES data set with signal and systematic models using the evidence to determine which models are preferred by the data. A further brief discussion of the evidence and Bayes theorem can be found in \cref{app:reproducability}.

Of the tested combinations of model components we find that modelling with a STA signal and no systematic leads to only a marginal increase in evidence in comparison to a foreground only fit as can be seen in the bottom panel of \cref{fig:model_fits_graph}. The increase in evidence is larger when we model an ERB signal but this is to be expected as these signals can have significantly deeper absorption features than the STA signals and as a result they are better able to fit out the larger systematic structure. Fits with systematic modelling have significantly higher evidences, regardless of whether we include a STA, ERB or no global signal model, than those without systematic modelling, $\Delta \log(Z) \approx 200 - 300$. The data, therefore, favours the presence of a systematic model but there is no strong indication for the presence of a signal in the data. 

\Cref{app:residuals} shows the residuals found when fitting the data with the PSF foreground model, the efficiency systematic and the constant noise model~(fit number 4) compared with the residuals from a high order polynomial fit. The consistency between the two sets of residuals suggests that the complexity of the modelling used here is sufficient to describe the data. Although a signal may still be present with an absolute maximum magnitude less than the noise after multiplication by the total efficiency of the antenna.

In the following sections we discuss in more detail the results found when modelling with the different components outlined in \cref{sec:modelling}.

\subsection{Noise}
\label{sec:noise_results}

\begin{figure}
    \centering
    \includegraphics{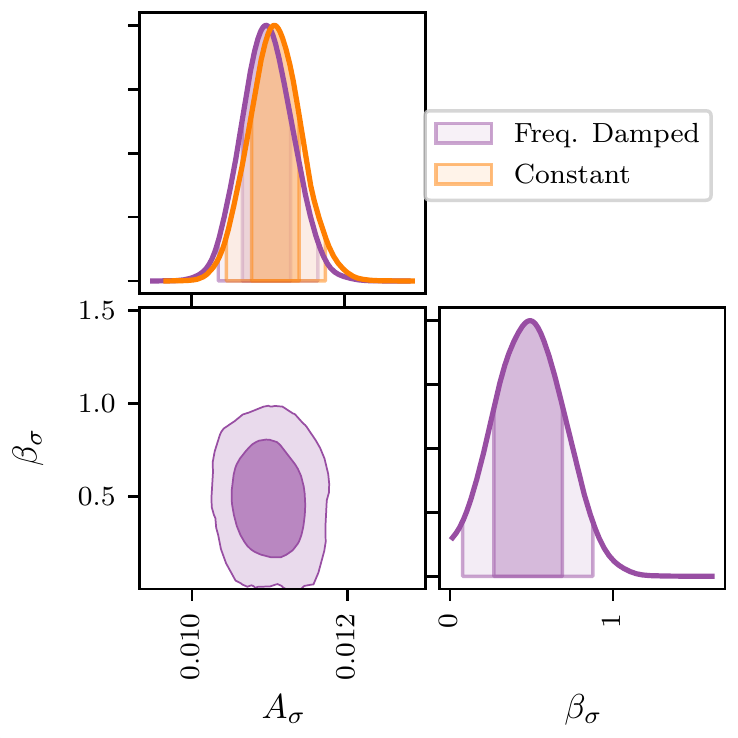}
    \caption{The posteriors for the noise parameters when fitting the SARAS2 data with STA signals and the efficiency systematic. The posteriors for the amplitude of the noise are near identical for the frequency damped (purple) and constant noise models (orange). In addition, the frequency damping power is approximately 0.5 which corresponds to very weak damping. In combination these two facts indicate a similarity between the two noise models. All of the corner plots presented in this paper were produced using \textsc{anesthetic} \citep{Handley2019} and the contour lines here represent 68\% and 95\% confidence intervals.}
    \label{fig:noise_posterior_example}
\end{figure}

We would expect that the relative weights based noise model would provide the best representation of the noise in the data as it has been derived from system parameters. In the absence of systematic modelling this noise model performs comparatively well because the residuals after foreground modelling are larger at higher frequencies following the trend of the weights. However, from the top panel of \cref{fig:model_fits_graph} it is clear that the log-evidence for fits with this noise model and a systematic is much lower than for fits with the two alternatives. This is likely due to some degeneracy, that would be hard to disentangle, between the systematic model and the weights based noise model which follow the same trend in the data (increasing with frequency). However, the systematic model, particularly the damped systematic, is designed to allow for no damping (i.e $\alpha_\mathrm{sys} = 0$ in \cref{eq:eff_sys} and \cref{eq:damped_sys}) and we find that some damping is typically favoured in our fits as is the presence of a systematic model.

For some combinations of signal and systematic, we find that the log-evidence is comparable between the fits with constant and frequency damped standard deviation. However, in the majority of cases the constant noise modelling gives a higher evidence. In the one case where this is not true, fitting with a STA signal and the efficiency systematic, the evidences are comparable indicating that neither noise model is favoured over the other.

For the fit with the efficiency systematic and STA signals we find, with the frequency damped noise, $\log(Z) = 1684.5 \pm 0.2$. In comparison, when modelling the noise with a constant standard deviation, we find $\log(Z) = 1684.4 \pm 0.2$. For these two fits, we can look in more detail at the maximum likelihood noise models and the posteriors for the noise model parameters are shown in \cref{fig:noise_posterior_example}. For the frequency damped model the maximum likelihood parameters are $A_\sigma = 10.9$ mK and $\beta_\sigma = 0.45$ and for the constant noise model the maximum likelihood amplitude is $A_\sigma = 11.0$ mK. In the former, the damping power is weak, recalling that the prior range for this parameter is $0 - 5$, and the maximum likelihood amplitude for both models is near identical with a similar posterior distribution. The weak damping is a consistent feature across all of the fits with the damped model as is the similarity in the amplitude of the noise.

To summarise, when comparing the two best noise models, constant and frequency damped, we find an equality in log-evidence or preference for the former model, a consistent weak damping in the later and similarity in amplitude. This supports the expectation from the literature that the noise should be uniform across the band in the SARAS2 data \citep{Singh2017,Singh2018}.

\subsection{Systematics}

As discussed, we see a large increase in log-evidence, $\Delta~\log(Z)~\approx 300$, when joint fits with a PSF foreground and the efficiency or damped systematic models are performed as shown in \cref{fig:model_fits_graph}. Comparing like for like fits with different systematic models we consistently find that the efficiency systematic model is favoured over the alternative by the data. Regardless of noise modelling, the five highest evidence models are all fits with the efficiency systematic.

Both systematics rely on the same parameterisation and so a direct comparison can be made between fits with the same noise and signal modelling but different systematic models. The posterior distributions for each parameter are typically well constrained Gaussian-like distributions. $\alpha_\mathrm{sys}$ is generally centered around 0 for the efficiency systematic in comparison to a value of approximately 3 for the damped systematic. This indicates that, in the case of the efficiency systematic, the data favours a sinusoidal systematic structure that is damped predominantly by the total efficiency of the radiometer, $\eta_t$ (i.e. the term $(\nu / \nu_0)^{\alpha_\mathrm{sys}} \approx 1$ in \cref{eq:eff_sys}).

For both systematic models the period, $P$, and phase, $\phi$, have similar distributions suggesting that any systematic in the data has a period of approximately $32.5$ MHz and phase of approximately $0$ rad. This is true generally, regardless of noise and signal model, since the systematic dominates the residuals. The systematics also have similar amplitudes but, while in the efficiency systematic this is mainly determined by $A$, in the damped systematic it is mainly determined by the term $(\nu / \nu_0)^{\alpha_\mathrm{sys}}$ in \cref{eq:damped_sys}.

From the log-evidences we can conclude that the data marginally favours a sinusoidal systematic damped by the total efficiency of the antenna. However, as with the noise modelling, the differences in log-evidence between like for like fits with different systematics are small and all of the fits with the different systematic models contain potential information about the astrophysical parameter space. We can, therefore, use all of the samples from the different fits, weighted by their evidences, in a combined analysis. This approach effectively marginalises over the systematic and noise model parameters and is detailed in the following section.

\subsection{Disfavoured 21-cm signals from combined samples}
\label{sec:disfavoured_regions_combined_samples}

We can use \textsc{anesthetic} \citep{Handley2019} to combine the posterior samples from \textsc{polychord}, weighted by the fit evidence, from the various nested sampling runs to determine the types of global 21-cm signals disfavoured in our analysis. This is advantageous since it provides a method to deal with uncertainty in the modelling of the standard deviation on the noise and the systematic and should be considered a conservative view of any constraints on the parameter space from the SARAS2 data.

While we note that the fits in \cref{fig:model_fits_graph} do not indicate a preference for the presence of a signal, we can still use the data to determine constraints on the parameter space of the global 21-cm signal as has previously been done with data from EDGES \citep{Monsalve2019}.

In the previous two sections we have made the argument that the amplitude of the noise is best described by a constant standard deviation and the systematic is best modelled by the efficiency systematic. While these statements are true to an extent the range in log-evidence between the corresponding fits with different model components is not significant. In fact, the former is largely motivated by the fact that the constant noise model is simpler than the frequency damped model and thus favourable. In practice the data does not tell us which noise model is preferred.

Since the 2D and 1D posteriors for the astrophysical parameters by definition marginalise over the systematic and noise parameters we can confidently combine the samples and draw conclusions from the corresponding posteriors. As alluded to the posterior samples, $P(\theta|D, M)$, when combined using \textsc{anesthetic} are weighted by weights, $w$, that are directly proportional to the fit evidence, $Z$
\begin{equation}
    P_\mathrm{combined}(\theta|D, M) = \sum_i w_i P_i(\theta|D, M),
    \label{eq:combined_samples}
\end{equation}
where the weights $w_i = Z_i/\sum_j Z_j$, $\theta$ is a vector of parameters associated with the fit components for fit $i$, $D$ is the data and $M$ is the analytical model. In \cref{app:relative_weights} we show the values of $w_i$ for each of the different fits performed in this analysis separated by their signal type.

In the following subsections we therefore discuss posteriors from combined samples for all of the fits containing STA signals (see \cref{app:lyman_res_vra} for results with variable values of $R_\mathrm{mfp}$ and $\alpha$) and fits containing ERB signals.

We briefly compare our results with those previously reported in \cite{Singh2017, Singh2018} and with the recent results from HERA in \cref{sec:comparison}.

\subsubsection{Excess Radio Background Signals}

\begin{figure*}
    \centering
    \includegraphics{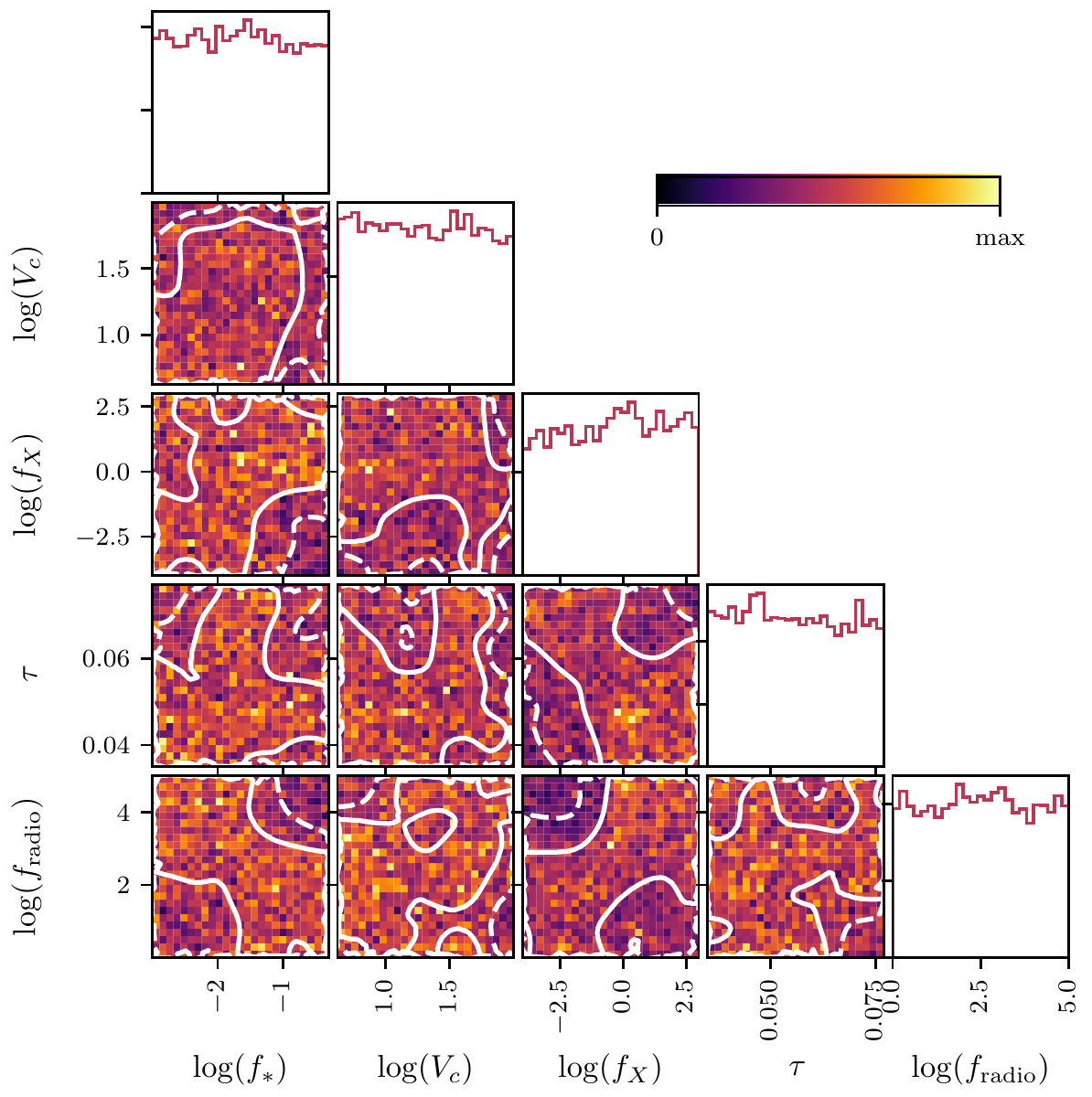}
    \caption{The 1D and 2D distributions for the combined nested samples of all fits to the SARAS2 data containing ERB signals. Again the solid contours in the 2D parameter space enclose 68\% confidence regions and the dashed contours enclose 95\% confidence regions. Combinations of high $f_*$ ($\gtrsim$ 0.03) and high $f_\mathrm{radio}$ ($\gtrsim$ 707) as well as low $f_X$ ($\lesssim$0.21) and high $f_\mathrm{radio}$ ($\gtrsim$ 407) are less densely sampled than the rest of the parameter space and are consequently disfavoured in our analysis.}
    \label{fig:combined_samples_fradio}
\end{figure*}

We can analyse combined samples from fit numbers 1, 5, 10, 12, 14 and 16, those containing ERB signals, as shown in \cref{fig:combined_samples_fradio}. The recovered 1D histograms are generally flat and do not show any significant constraints. Although regions of high $f_\mathrm{radio}$ ($\gtrsim$ 407 and $\gtrsim$ 707)
in combination, separately, with low $f_X$ ($\lesssim$ 0.21) and high $f_*$ ($\gtrsim$ 0.03) are disfavoured at 68\% confidence. We also disfavour high values of $\tau$ above approximately $\gtrsim$ 0.06 in combination with $f_X \gtrsim 0.50$ at 68\% confidence.

We use histograms to illustrate the samples in the parameter space rather than plotting posteriors, with confidence regions, derived using Kernel Density Estimation~(KDE) as was done in \cref{fig:noise_posterior_example}. This is done because the application of a KDE to the samples can, in the case of flat distributions, lead to misleading features that suggest specific areas of the parameter space are more favourable than others. Binning the raw samples across the parameter space gives a clearer impression of regions of the parameter space that the nested sampling algorithm explored in greater detail (i.e. those with higher likelihoods and combinations of parameters that are favoured by the data). For example in \cref{fig:combined_samples_fradio} favoured regions of the parameter space would be sampled in greater detail and the corresponding bins in the 2D plots would be shaded in a lighter yellow.

To quantify the strength of 2D constraints across the parameter space the authors of \cite{HERA} assess the ratio of the minimum and maximum posterior probability ($\propto$ height of the histogram bins) across the 2D space. A ratio of 1 would indicate a perfectly flat posterior and a low value indicates a non-uniform distribution. The metric is limited in that it does not indicate the direction of any non-uniformity and is insensitive to any pseudo-random scatter across the parameter space that may result from a fine binning of a relatively flat distribution\footnote{The metric is also dependent on the number of bins into which the samples are separated and so comparison across experiments is difficult without fixing the number of bins in each set of analysis.}. However, in combination with a visual inspection of a constrained parameter space it can be useful metric to determine the magnitude of any directional non-uniformity. For example, the ratio between the minimum and maximum posterior probability for $f_\mathrm{radio}-f_X$ in \cref{fig:combined_samples_fradio} is 0.10. In comparison the maximum ratio for \cref{fig:combined_samples_fradio} is 0.20 for $V_c - f_\mathrm{radio}$ and the minimum ratio is 0.08 for $f_X - \tau$.

\begin{figure*}
    \centering
    \includegraphics{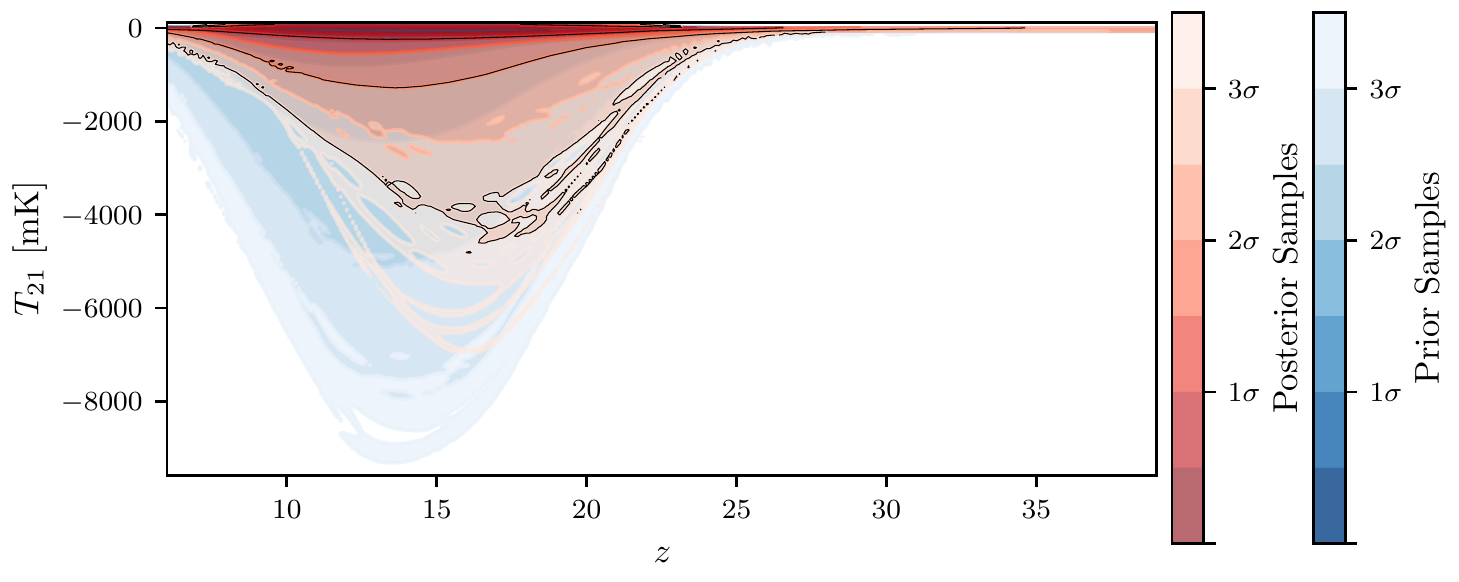}
    \caption{The figure shows the contraction from the prior (blue) to the combined posterior samples (red) in the 5D astrophysical parameter space (i.e. $T_{21} - z$) for the ERB signals. We plot the prior and posterior on top of each other. The figure was produced using \textsc{fgivenx} \protect\citep{fgivenx} and indicates that we rule out the deepest of these exotic astrophysical models.}
    \label{fig:fradio_fgivenx}
\end{figure*}

Further, the deepest signals have high Lyman-$\alpha$ fluxes (high $f_*$ and low $V_c$), low X-ray efficiencies, $f_X$, and high values of $f_\mathrm{radio}$. After multiplication by the total efficiency of the antenna these signals typically have magnitudes larger than the expected noise and consequently we would expect to exclude these with the SARAS2 data. This can be seen clearly in \cref{fig:fradio_fgivenx} in which we plot the functional posterior samples on top of the prior using the tool \textsc{fgivenx} \citep{fgivenx}. The tool allows us to visualize the combined posterior samples shown in \cref{fig:combined_samples_lyh_fra} as a set of contours in the $T_{21} - z$ space. Note that although the priors on our parameters are uniform this does not necessarily translate to a uniform prior in the global 21-cm signal.

\subsubsection{Standard Astrophysical Signals}

\begin{figure*}
    \centering
    \includegraphics{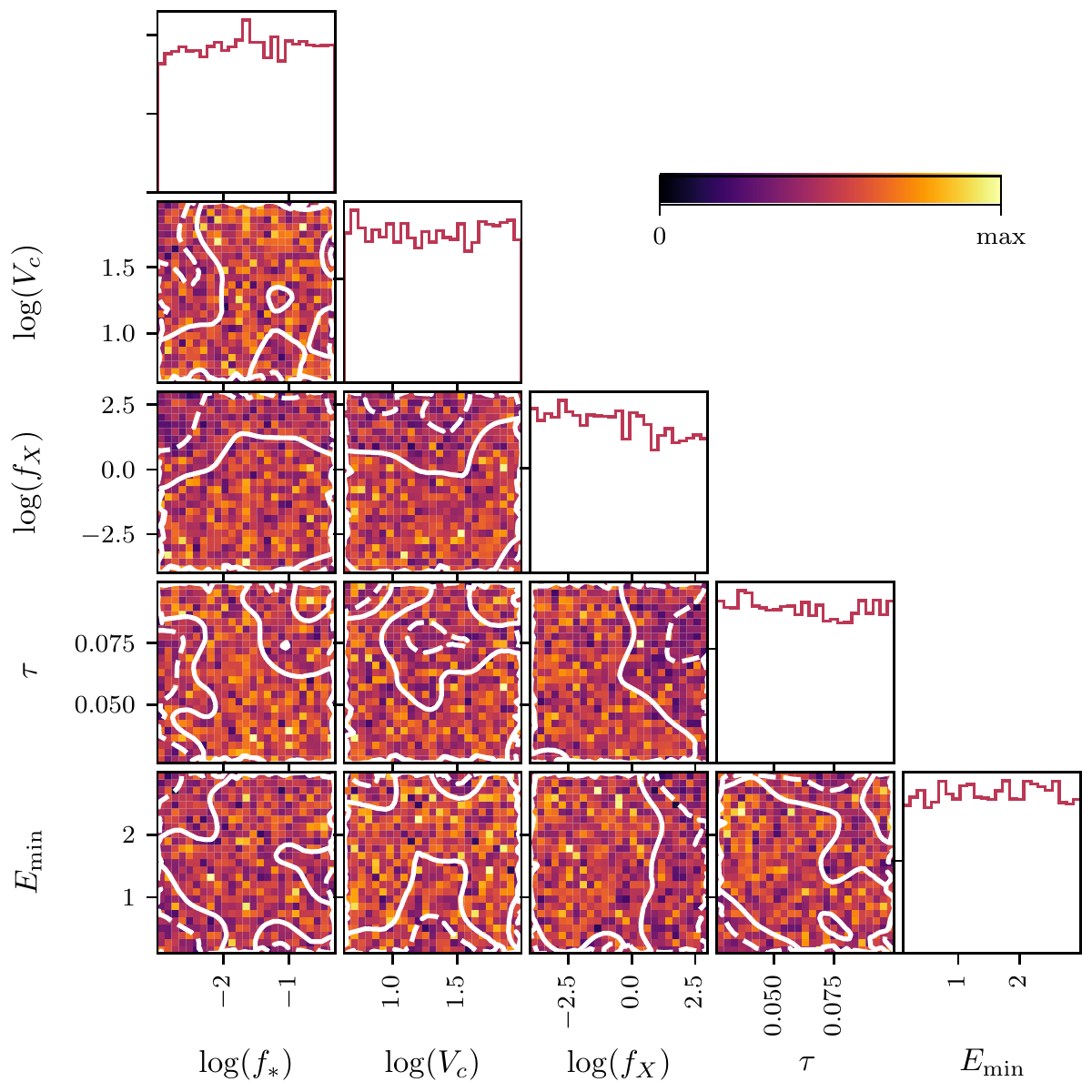}
    \caption{The 1D and 2D histogram plots from the combined nested samples of all fits to the SARAS2 data containing STA signals with fixed values of $R_\mathrm{mfp}$ and $\alpha$. The solid contours in the 2D posteriors enclose regions of 68\% confidence and the dashed contours enclose regions of 95\% confidence.
    The histograms are relatively flat and we do not disfavour values of individual parameters.}
    \label{fig:combined_samples_lyh_fra}
\end{figure*}

Histograms produced from the combined samples for the astrophysical parameters from fit numbers 2, 3, 9, 11, 13, and 18 from \cref{fig:model_fits_graph}, those with STA signals, are shown in \cref{fig:combined_samples_lyh_fra}. The histograms are flat indicating that we do not significantly constrain the parameter space and further indicating that there is no preference for a signal in the data.

\begin{figure*}
    \centering
    \includegraphics{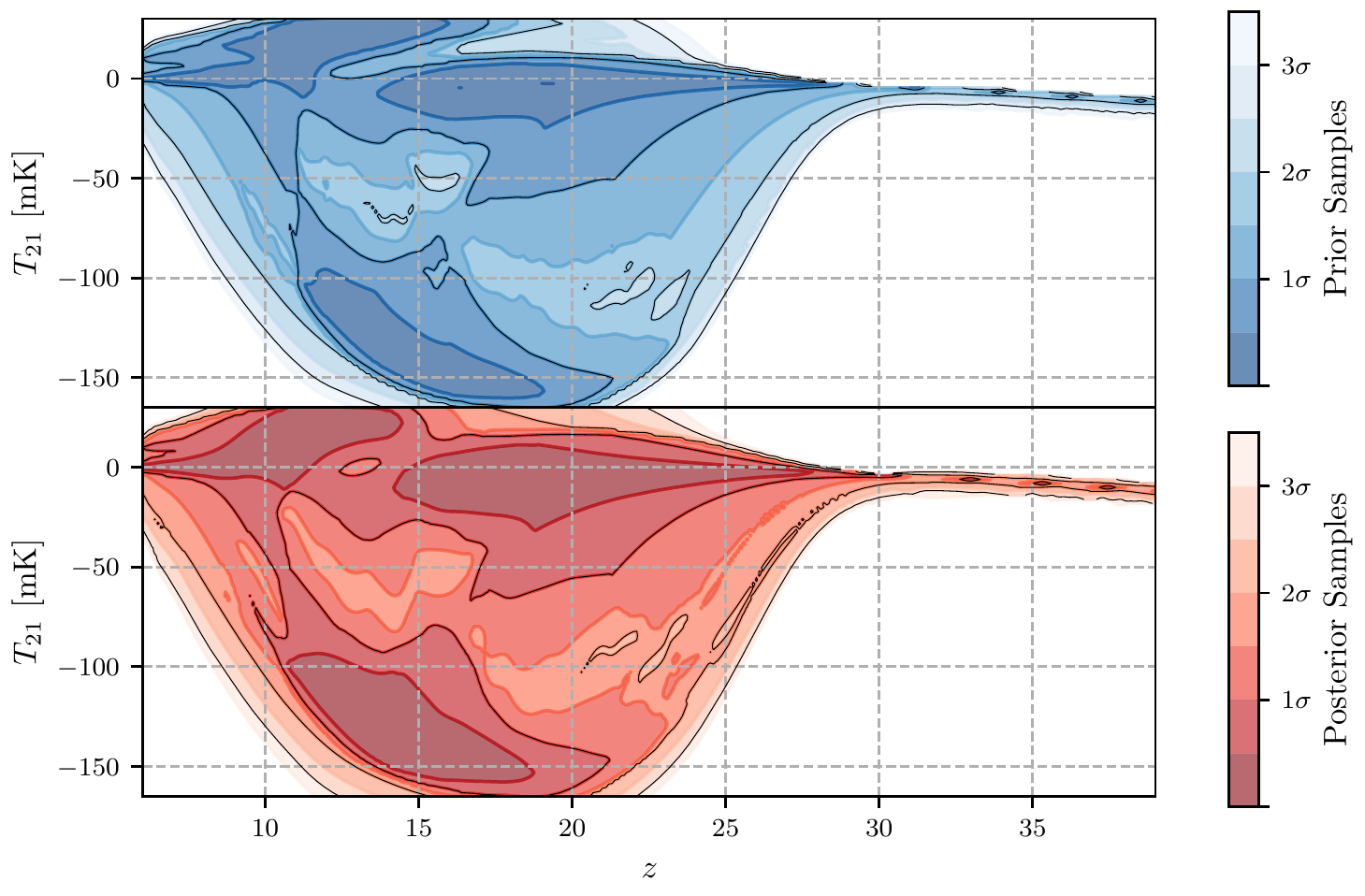}
    \caption{The figure shows the contraction from the prior (blue) to the combined posterior samples (red) in the 5D astrophysical parameter space (i.e. $T_{21} - z$) for the STA signals. The graph indicates that the data weakly disfavours (lighter shaded red regions) standard astrophysical signals with absorption features at high redshift. However, we note that the contraction from the prior to the posterior is minimal. Again the figure was produced using \textsc{fgivenx} \protect\citep{fgivenx} and we have added gridlines to make the differences between the two distributions clearer.}
    \label{fig:fgivenx_lah}
\end{figure*}

\cref{fig:fgivenx_lah} shows the functional posterior (in red) along with the equivalent for samples taken from the prior (in blue). There is some indication that the data disfavours (lighter shaded red regions) signals with absorption features at high redshift. For example, the preferred region (darker red and blue with significance $<1\sigma$) around the absorption minimum, where the sampling is highest, is larger and shifted to lower redshifts in the posterior than in the prior. However, the contraction from the prior to the posterior is minimal as would be expected from the flat nature of the distributions in \cref{fig:combined_samples_lyh_fra}. Therefore any conclusions we make about constraints on the standard astrophysical priors are by definition weak. 

The contraction from prior to posterior can be quanitified with the Kullback-Leibler Divergence and Bayesian Dimensionality \citep{handley_dimensionality_2019}. However, we are only interested in the contraction from specifically the astrophysical prior to the astrophysical posterior and we do not want to include contributions to the statistics from nuisance parameters, like the systematic parameters, in our calculations. This requirement makes quantifying the contraction for the analysis performed here non-trivial. We therefore leave a detailed discussion of these marginal Bayesian statistics to future work (Bevins et al. in prep).

\subsubsection{Results in the context of the previous SARAS2 analysis and the HERA results}
\label{sec:comparison}

As discussed in \cref{sec:intro}, previous analysis of the SARAS2 data, with standard astrophysical signal models, led to the conclusion that a set of specific astrophysical models with rapid reionization and weak X-ray heating (low values of $f_X$) were disfavoured. We find that we weakly disfavour standard astrophysical signals with high Lyman-$\alpha$ fluxes and weak heating and this can be seen in \cref{fig:fgivenx_lah}. Our constraints are stronger on the ERB signals in which we also find that we disfavour signals with weak X-ray heating, in particular, although these signals were not explored in the previous SARAS2 analysis. It is important to note that the simulations used in this paper have different parameterizations of the X-ray SEDs and the original SARAS2 analysis was confined to the study of a limited sample of 264 models. 

Further, the signals used in each study have significantly different dependence on the Lyman-$\alpha$ flux with the introduction of Lyman-$\alpha$ heating in the simulations used in this work. Additional heating from the CMB is also included here which will influence the conclusions we make about the strength of the X-ray heating and makes comparison between the two sets of analysis difficult. 

The additional heating mechanisms influence the position of the minima of the global signal with a higher heating shifting the minima to higher redshifts outside the SARAS2 band and consequently reducing the magnitude of the signal at low redshift. This in turn leads to a lower signal to noise ratio in the data and makes signals with high heating hard to rule out. In contrast signals with weak heating have minima at lower redshifts, have larger absolute magnitudes in the SARAS2 band and higher signal to noise ratios making them easier to rule out.

Lyman-$\alpha$ and CMB heating, as discussed in \cref{sec:signal_modelling}, actually reduce the maximum allowed depth of the global signal to approximately $\approx 165$ mK in comparison to the predictions from the previous state of the art simulations, used in the original SARAS2 analysis, which could reach depths of $250$ mK \citep{Cohen2017}. It is this difference in the maximum absolute magnitude between the two sets of signals used in each set of analysis which explains why the previous analysis is able to rule out a set of signals with such high confidence ($> 5\sigma$ in some cases) and in this paper we are unable to make any significant conclusions for standard astrophysical signals. The majority of the signals ruled out in the previous work have amplitudes $\gtrsim 125$ mK and so if present in the SARAS2 data would have a higher signal to noise ratio than the majority of the models analysed in this paper with amplitudes $ \lesssim 165$ mK making them easier to identify or rule out.

Comparison with results from other experiments in the literature is generally difficult because the analysis often uses contemporary parameterizations of the global signal that are subsequently superseded by more astrophysically accurate models. However, the recent upper limits on the 21-cm power spectrum at $z\approx8$ and 10 by the HERA~(Hydrogen Epoch of Reionization Array) collaboration have been used to derive parameter constraints on an excess radio background from high redshift radio galaxies using the same parameterization used in this paper across the band $z \approx 7 - 12$ \citep{HERA}.

As discussed, SARAS2 data is most sensitive to and able to exclude the ERB signals with the largest radio backgrounds because their magnitudes within the SARAS2 band are larger than the experimental noise. In a similar way, the authors in \cite{HERA} are able to exclude models with high radio backgrounds because the corresponding power spectra are larger than the upper limits provided by HERA.

The authors perform parameter constraints using the likelihood function described in their section~(3), the MCMC Ensemble sampler \textsc{emcee} \citep{emcee} and a neural network emulator of the power spectrum detailed in their appendix A. When investigating the parameter constraints on the ERB models the authors rule out, with a higher significance, a similar region of the $f_\mathrm{radio} - f_X$\footnote{Note that $f_\mathrm{radio}$ is referred to by $f_r$ in \cite{HERA}.} parameter space as is done in this paper with the SARAS2 data. Specifically the authors rule out values of $f_X < 0.33$ and $f_\mathrm{radio} > 391$ in comparison to our values of 0.21 and 407 respectively.

\section{Conclusions}
\label{sec:conclusions}

In this paper we have reported constraints on the EoR using data from the SARAS2 experiment, the nested sampling algorithm \textsc{polychord}, the derivative constrained function fitting code \textsc{maxsmooth} and the global signal emulator \textsc{globalemu}. 

We have fitted, to data from a global 21-cm experiment for the first time, standard astrophysical signals with Lyman-$\alpha$ and CMB heating and exotic astrophysical models with an excess radio background produced from high redshift galaxies. General conclusions from our analysis are summarised below;

\begin{itemize}
    \item We have found no conclusive evidence for the presence of a signal in the data and fits performed with and without signal modelling have comparable evidences.
    \item The data generally favours the presence of noise with a constant standard deviation across the SARAS2 band over the two alternatives tested in this paper.
    \item We have illustrated the presence of a damped sinusoidal systematic in the data using the smooth foreground model implemented with \textsc{maxsmooth}. Our analysis suggests that this systematic is best modelled as a sinusoidal function that has been damped by the total efficiency of the antenna. This implies that the systematic is introduced as power external to the radiometer rather than via the electronics in the receiver chain and back-end. However, we note that the log-evidence difference between the fits performed with the two different systematic models is marginal and that the non-smooth structure may have been introduced by a poor foreground model. If real, the systematic could be caused by discontinuities in the soil surrounding the antenna or shrubbery and root systems in close proximity both of which are issues that will be alleviated by the deployment of SARAS3 on a lake.
    \item While we do not constrain individual parameters, for the ERB signals, we disfavour combinations of high $f_*$ ($\gtrsim$ 0.03 and high $f_\mathrm{radio}$ ($\gtrsim$ 707) and low $f_X$ ($\lesssim$0.21) and high $f_\mathrm{radio}$ ($\gtrsim$ 407) which produce the deepest absorption troughs as can be seen in \cref{fig:combined_samples_fradio} and \cref{fig:fradio_fgivenx}. In addition, when fitting the ERB signals we weakly disfavour high values of $\tau$ and the combination of low $\tau$ and low $f_X$.
    \item For standard astrophysical models we weakly disfavour signals with high Lyman-$\alpha$ fluxes (high $f_*$ and low $V_c$) and weak heating that have deep absorption features at early times as can be seen in \cref{fig:fgivenx_lah} (with the aid of the gridlines). 
    \item Both sets of analysis with STA and ERB signals disfavour models with weak heating, particularly X-ray heating in latter case, in agreement with the SARAS2 data analysis in \cite{Singh2017} and \cite{Singh2018}.
    \item We disfavour a similar combination of low $f_X$ and high $f_\mathrm{radio}$ for the ERB models as was recently done using the power spectrum upper limits from HERA with an identically parameterized model of the EoR \citep{HERA}.
\end{itemize}

The analysis presented here serves to highlight that non-smooth systematics, if effectively identified with tools like \textsc{maxsmooth} and modelled, do not prevent the derivation of constraints on the astrophysics of the early universe.

\section*{Acknowledgements}

HB acknowledges the support of the Science and Technology Facilities Council (STFC) through grant number ST/T505997/1. WH and AF were supported by Royal Society University Research Fellowships. EA was supported by the STFC through the Square Kilometer Array grant G100521. RB acknowledges the support of the Israel Science Foundation (grant
No. 2359/20), The Ambrose Monell Foundation and the Institute for
Advanced Study.

\section*{Data Availability}

The SARAS2 data was provided by the SARAS2 collaboration and is not publicly available. The global 21-cm signal models were provided by A. Fialkov and collaborators and are not publicly available.

%%%%%%%%%%%%%%%%%%%%%%%%%%%%%%%%%%%%%%%%%%%%%%%%%%



\bibliographystyle{mnras}
\bibliography{saras}

%%%%%%%%%%%%%%%%%%%%%%%%%%%%%%%%%%%%%%%%%%%%%%%%%%

%%%%%%%%%%%%%%%%% APPENDICES %%%%%%%%%%%%%%%%%%%%%

%%%%%%%%%%%%%%%%%%%%%%%%%%%%%%%%%%%%%%%%%%%%%%%%%%

\appendix

\section{Reproducability of Results}
\label{app:reproducability}

The nested sampling algorithm used in this work is designed to numerically approximate the integral
\begin{equation}
    P(D|M) = \int P(D|\theta, M) P(\theta|M) d\theta,
    \label{eq:bayes_integral}
\end{equation}
or equivalently
\begin{equation}
    Z = \int \mathcal{L}(\theta) \pi(\theta) d\theta,
\end{equation}
where $Z = P(D|M)$ is known as the evidence, $\mathcal{L}(\theta) = P(D|\theta, M)$ is the likelihood and $\pi(\theta) = P(\theta|M)$ is the prior probability. The evidence can be used to determine whether one model is a better description of the data than another (i.e. model selection) as is done in the main text in this paper. The prior represents our knowledge of the parameters in our model and typically is taken to be a uniform or log-uniform probability distribution between a minimum and maximum value. Finally, the likelihood represents the probability that we observe the data, $D$, given the choice of parameters and model or hypothesis, $M$, to describe the data. A complete discussion of the algorithm can be found in \cite{skilling_nested_2004}.

\Cref{eq:bayes_integral} can be derived from Bayes' theorem
\begin{equation}
    P(\theta|D, M) = \frac{\mathcal{L}(\theta) \pi(\theta)}{Z},
\end{equation}
where $P(\theta |D, M)$ is the posterior probability used to determine constraints in the parameters $\theta$, and the requirement that the posterior should integrate to 1. The posterior is therefore a byproduct of the nested sampling algorithm and its accuracy is determined by the accuracy of approximation of the integral in \cref{eq:bayes_integral}.

The accuracy of the integral, in turn, is determined by the number of likelihood samples taken when running the algorithm with tools like \textsc{polychord}. The sampling rate in \textsc{polychord} is driven by the parameter $n_\mathrm{live}$ and a poor sampling leads to poor reproducibility of the posteriors on repeated runs.

\begin{figure*}
    \centering
    \includegraphics{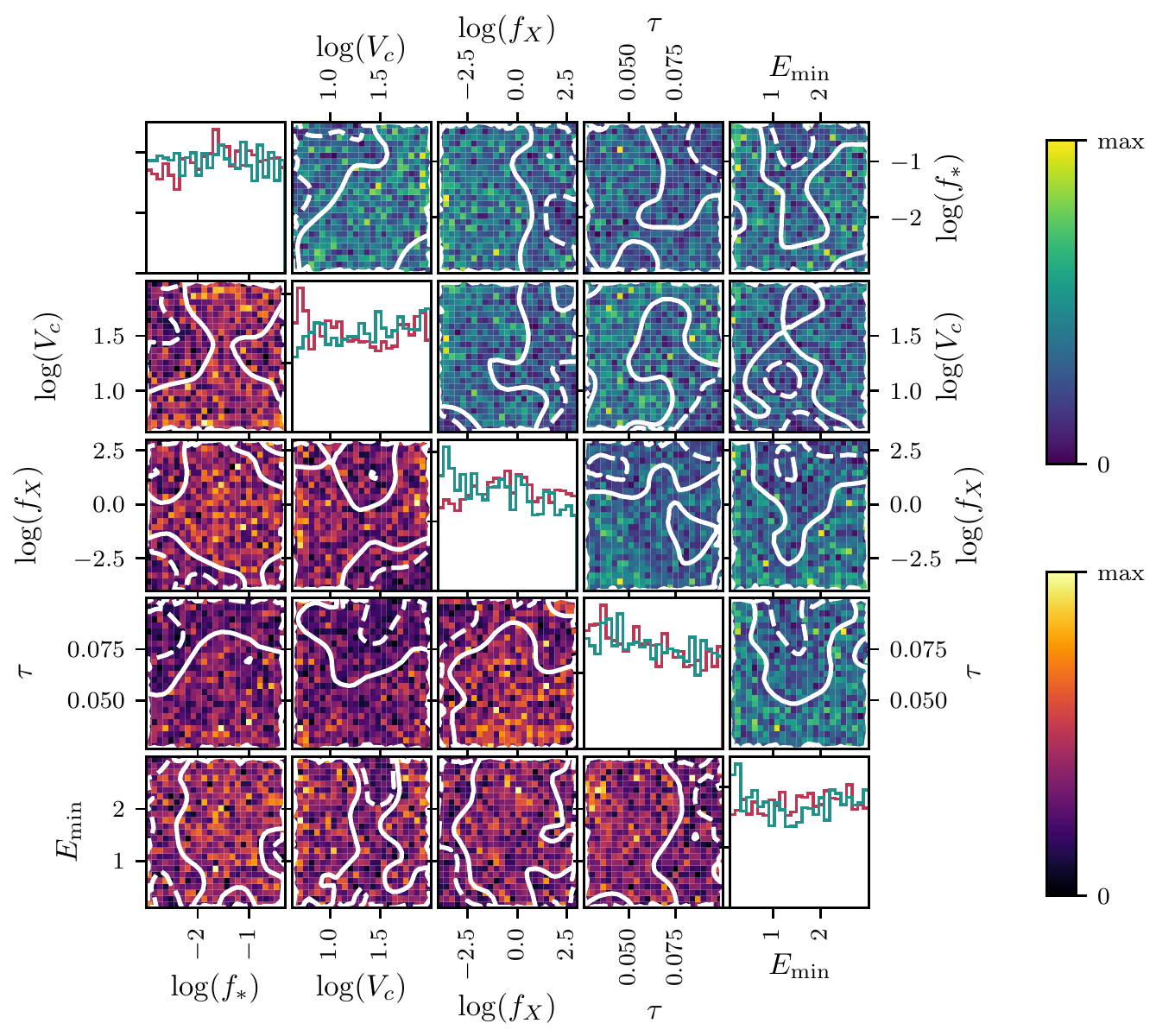}
    \caption{The sample distribution for fit number 3 in \cref{fig:model_fits_graph} (STA signal, efficiency systematic and constant noise) and a corresponding repeated run demonstrating that the posterior sampling is high enough to lead to reproducible results. The consistency between repeated runs shown here is typical for all of the fits performed in this paper.}
    \label{fig:reproducability}
\end{figure*}

For the analysis in this paper we use $n_\mathrm{live} = 500$ which equates to approximately 50 live points per dimension. We demonstrate that this leads to reproducible sample distributions in \cref{fig:reproducability} which shows histograms of the distributions for fit number 3 (STA signal, efficiency systematic and constant noise) and a corresponding repeated run. In both cases the recovered 1D histograms are flat with only minor differences. 

We can quantify the difference using the two sample Kolmogorov-Smirnov (KS) statistic which returns the maximum difference between two empirical cumulative distribution functions. The largest KS statistic for the 1D distributions in \cref{fig:reproducability} is $0.073$ for $\log(f_X)$. For all other parameters the KS statistic is smaller. For a given parameter a low KS statistic, which ranges in value between 0 and 1, indicates that the two 1D distributions are likely drawn from the same sample and consequently the results are reproducible.

\section{Model Complexity and the Expected Noise}
\label{app:residuals}

\begin{figure*}
    \centering
    \includegraphics{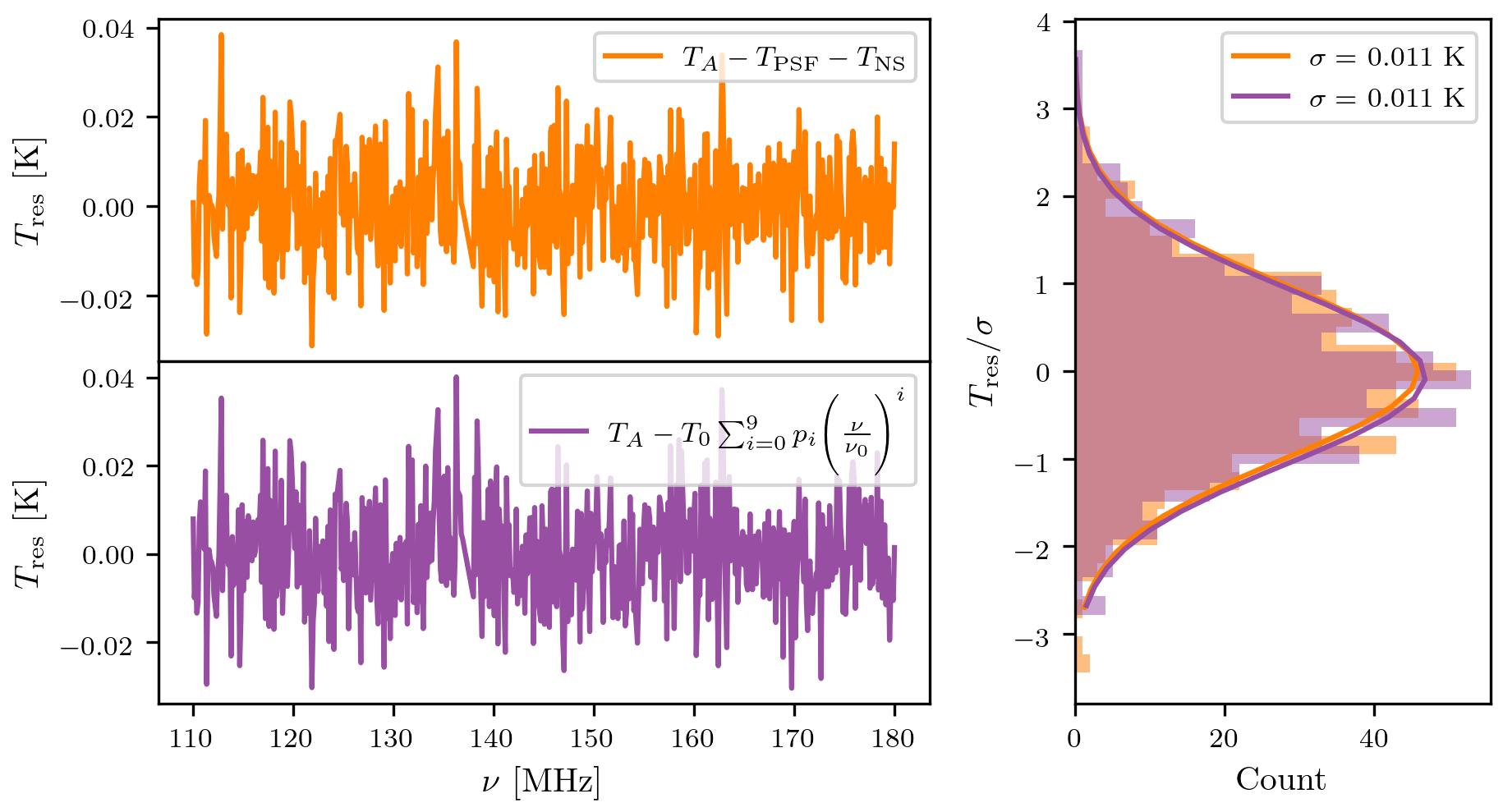}
    \caption{The figure shows the residuals found when fitting the SARAS2 data, $T_A$, with the PSF foreground, the efficiency systematic model and a constant noise model compared to the residuals from a high order polynomial fit. The unconstrained polynomial fit is expected to model out all non-smooth structure in the data including any systematics and signals revealing the noise in the data. The consistency between the two sets of residuals, which can be seen in the accompanying histogram, suggests that the complexity of our modelling (foreground, systematic and noise) is sufficient to describe the data. The graph also shows that the noise in the data is Gaussian distributed.}
    \label{fig:residuals}
\end{figure*}

\cref{fig:residuals} shows the residuals~(top left) after fitting and subtracting the PSF foreground model and the efficiency systematic model from the SARAS2 data compared with the residuals~(bottom left) from a high order polynomial fit given by
\begin{equation}
    T = T_0 \sum^9_{i=0} p_i \bigg(\frac{\nu}{\nu_0}\bigg)^i
\end{equation}
where $\nu_0 \approx 110$~MHz, $T_0 = T_A(\nu_0)$ and $p_i$ are the fitted coefficients. In the right hand panel we show a histogram of the two sets of residuals with corresponding Gaussian fits. The standard deviation from both fits are equivalent.

The high order unconstrained polynomial is expected to fit out any non-smooth structure in the data and as a result the residuals are expected to be representative of the noise. The graph therefore shows two things. Firstly, that our assumption that the noise in the data is Gaussian distributed holds. Secondly, that the complexity of our model~(foreground plus systematic) is sufficient to describe the SARAS2 data. Any signal in the data will have a maximum absolute magnitude less than the noise after multiplication by the total efficiency of the antenna and as a result the noise floor allows us to apply the constraints detailed in the text.

\section{Sample Weighting}
\label{app:relative_weights}

\begin{figure*}
    \centering
    \includegraphics{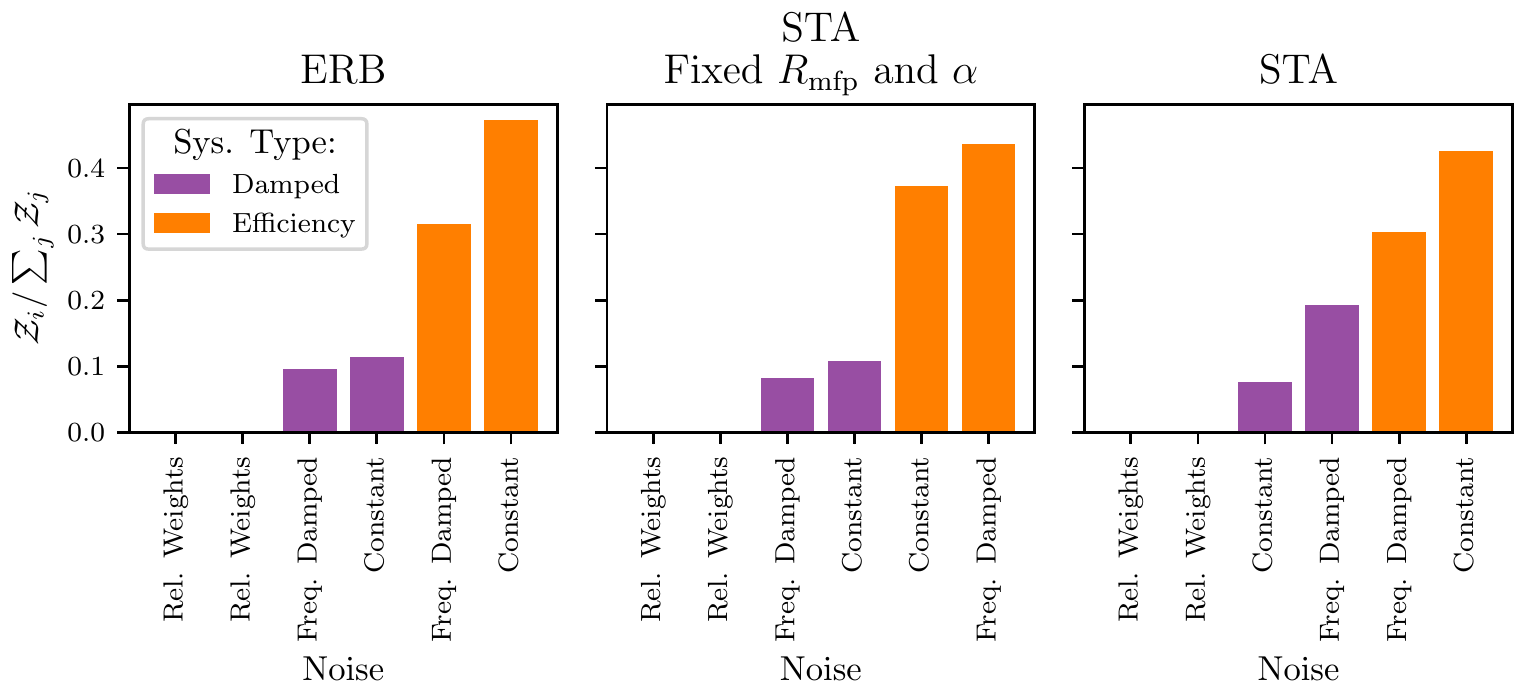}
    \caption{The figure shows the relative weights applied to each set of samples when combining the posterior distributions using \cref{eq:combined_samples}. The figure illustrates that the weights, which are based on the relative evidences of the fits, are significantly higher for fits performed with the Efficiency systematic and fits performed with the Constant or Frequency Damped noise model over the alternatives. By combining the samples to draw conclusions about the astrophysical parameter space we effectively account for any uncertainty in the systematic and noise modelling.}
    \label{fig:relative_weights}
\end{figure*}

Using \cref{eq:combined_samples} we are able to combine the samples with common astrophysical signals in order to marginalise out our uncertainty in the modelling of the noise and the systematic. To take account for the different levels of confidence in the different models, where the model refers to the combination of foreground, signal and noise model, we weight the samples by their Bayesian evidence. The weights are given by $Z_i/\sum_j Z_j$ and in \cref{fig:relative_weights} we show the weights for each signal type.

From the figure it is evident, regardless of signal modelling, that fits with the Efficiency systematic model have a higher weighting than fits with the Damped systematic indicating a preference for the former. Additionally, the figure shows that the samples from fits with the Relative Weights based noise are down weighted significantly so that they do not contribute to the calculation of the combined posteriors. We can see that this is the case by looking at the betting odds between two of the ERB fits both with the Efficiency systematic but with Constant and Relative Weights based noises. The difference in evidence for these two fits is given by $\exp(1684.5-1678.6) \approx 365$ which corresponds to betting odds of 365:1 in favour of the fit with a constant standard deviation on the noise.

\section{Disfavoured regions when fitting with standard astrophysical signals and variable $R_\mathrm{mfp}$ and $\alpha$}
\label{app:lyman_res_vra}

\begin{figure*}
    \centering
    \includegraphics{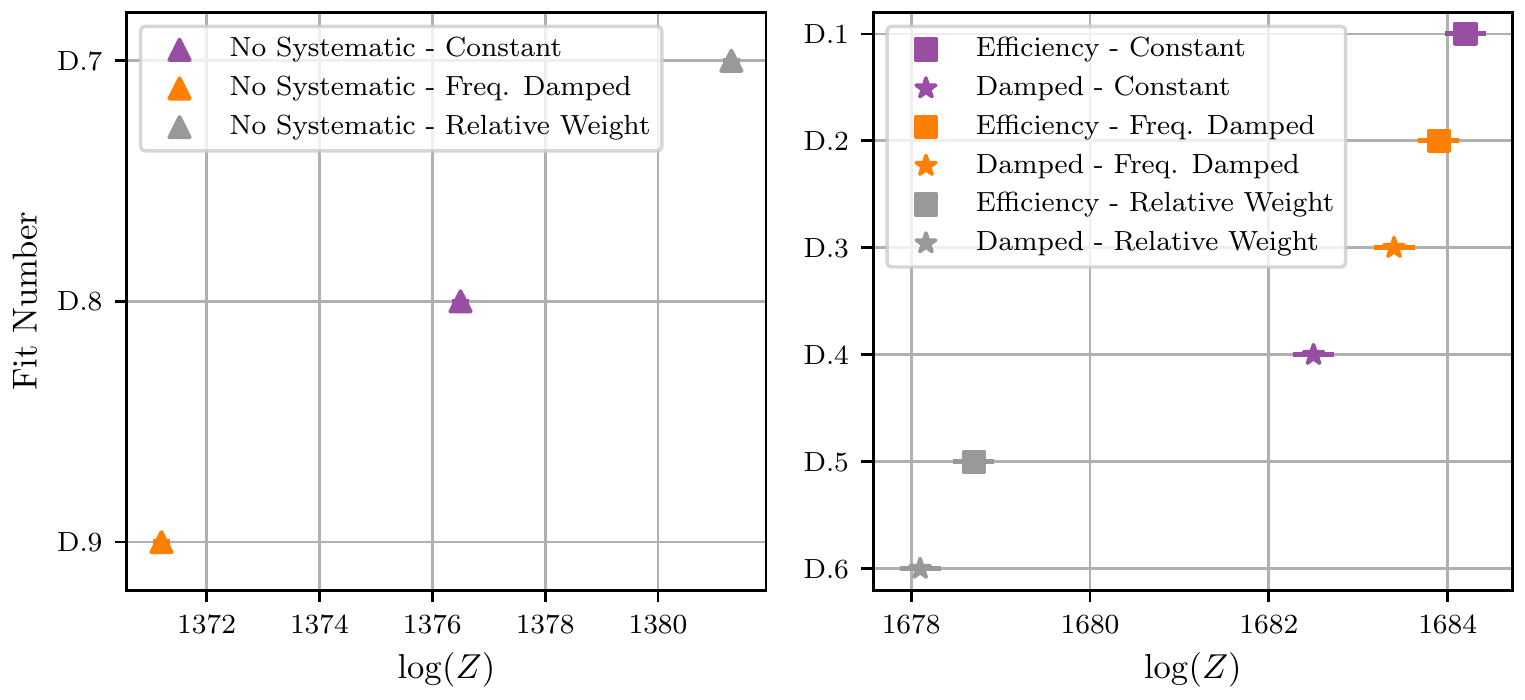}
    \caption{The evidence for fits containing STA signals when fitting for $R_\mathrm{mfp}$ and $\alpha$ rather than holding them constant. Allowing these parameters to vary has little effect on the overall patterns seen in preference for specific model components.}
    \label{fig:vra_only_evidences}
\end{figure*}

\begin{figure*}
    \centering
    \includegraphics{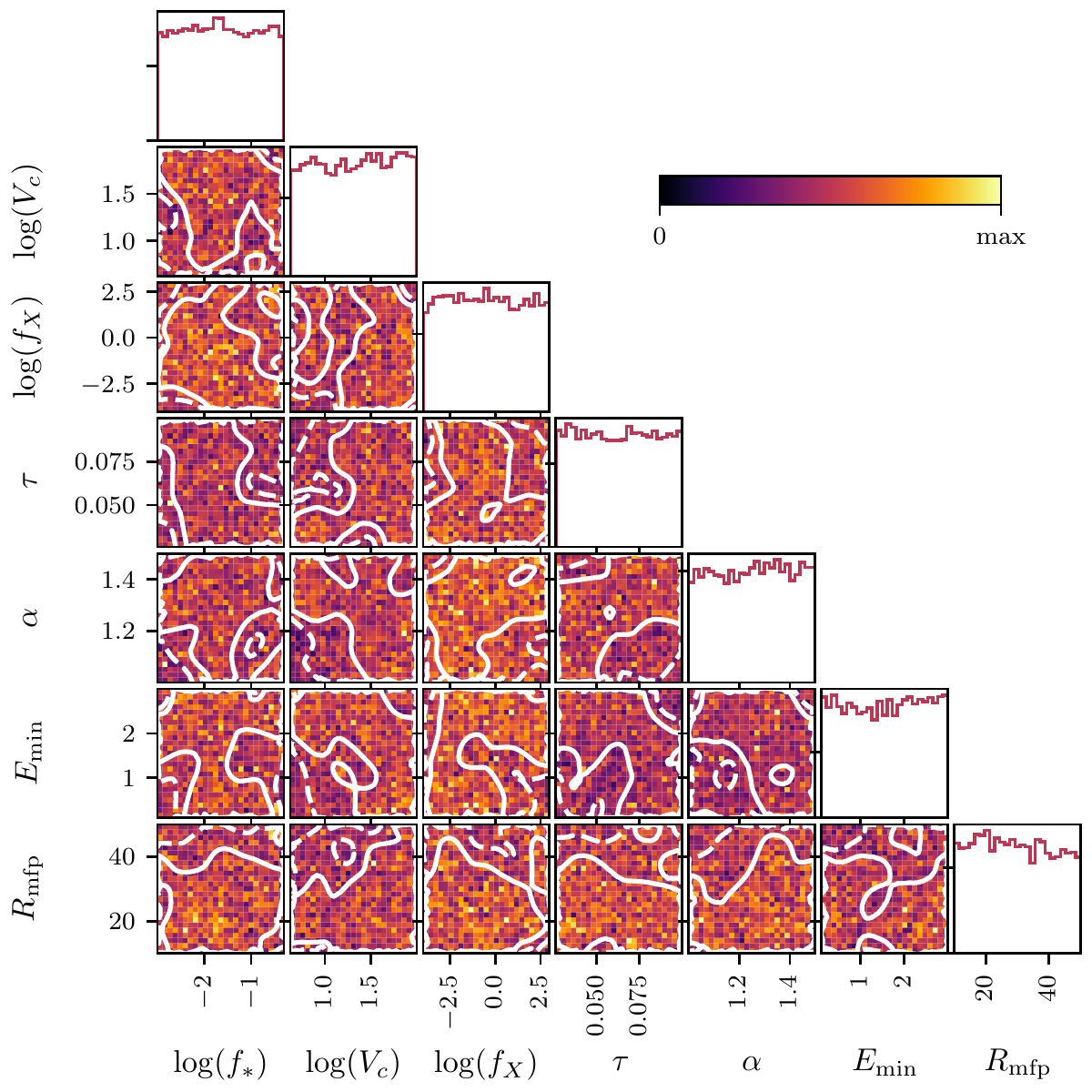}
    \caption{The 1D and 2D histograms for the combined nested samples of all fits to the SARAS2 data containing STA signals when fitting for $R_\mathrm{mfp}$ and $\alpha$.}
    \label{fig:combined_samples_lyh_vra}
\end{figure*}

For completeness we can assess combined samples from model numbers D.1, D.2, D.3, D.4, D.5 and D.6 in \cref{fig:vra_only_evidences}, those that contain STA signals with variable $R_\mathrm{mfp}$ and $\alpha$. \cref{fig:combined_samples_lyh_vra} shows the 1D histograms from the combined samples and, as with the results presented in the main text, we do not see any significant constraints.

However, the evidences presented in \cref{fig:vra_only_evidences} show a similar preference for different model components as that presented in the main text. Specifically, the preference for the efficiency systematic and for the frequency damped/constant noise models over the relative weights based noise.

% Don't change these lines
\bsp	% typesetting comment
\label{lastpage}
\end{document}